\documentclass[letterpaper,twocolumn,10pt]{article}
\usepackage{usenix2019_v3}

\usepackage{tikz}
\usepackage{amsmath}
\usepackage{amssymb}

\usepackage{xcolor}

\usepackage{algorithm}
\usepackage{algorithmic}

\usepackage[capitalize]{cleveref} 

\usepackage{xspace}
\usepackage{enumitem}

\usepackage{tcolorbox}

\usepackage{multicol}
\usepackage{multirow}
\usepackage{makecell}
\usepackage{booktabs}

\definecolor{olivegreen}{rgb}{0, 0.6, 0}
\definecolor{redorange}{HTML}{FF5349}
\definecolor{blue(ncs)}{rgb}{0.0, 0.53, 0.74}
\definecolor{navy}{HTML}{273BE2}

\definecolor{black}{HTML}{000000}
\definecolor{white}{HTML}{ffffff}
\definecolor{color1}{HTML}{ACE5EE}
\definecolor{color2}{HTML}{0093AF}
\definecolor{color3}{HTML}{CC0000}
\definecolor{color4}{HTML}{0087BD}
\definecolor{color5}{HTML}{333399}
\definecolor{color6}{HTML}{20B2AA}

\usepackage{xcolor}

\usepackage{soul}

\usepackage{marginnote}

\usepackage{pifont}

\newcommand{\thiswork}{PathWeaver\xspace}

\newcommand{\JL}[1]{{\color{cyan}[\textbf{\sc JLee}: \textit{#1}]}}
\newcommand{\JY}[1]{{\color{blue(ncs)}[\textbf{\sc JSong}: \textit{#1}]}}
\newcommand{\SY}[1]{{\color{violet}[\textbf{\sc SPark}: \textit{#1}]}}
\newcommand{\SJ}[1]{{\color{blue}[\textbf{\sc SKim}: \textit{#1}]}}
\definecolor{skyblue}{RGB}{30, 144, 255}
\newcommand{\JU}[1]{{\color{skyblue}[\textbf{\sc JHong}: \textit{#1}]}}
\newcommand{\MAPAE}[1]{{\color{orange}[\textbf{\sc HLim}: \textit{#1}]}}
\newcommand{\KTH}[1]{{\color{brown}[\textbf{\sc TKwon}: \textit{#1}]}}
\newcommand{\SN}[1]{{\color{teal}[\textbf{\sc SNoh}: \textit{#1}]}}
\newcommand{\SHP}[1]{{\color{orange}[\textbf{\sc Shepherd}: \textit{#1}]}}

\iftrue   
\renewcommand{\JL}[1]{}
\renewcommand{\JY}[1]{}
\renewcommand{\SY}[1]{}
\renewcommand{\JU}[1]{}
\renewcommand{\MAPAE}[1]{}
\renewcommand{\KTH}[1]{}
\renewcommand{\SN}[1]{}
\else
\fi

\iffalse  
\renewcommand{\SHP}[1]{}
\renewcommand{\SJ}[1]{}
\else
\fi

\newcommand{\naive}{naïve\xspace}
\newcommand{\Naive}{Naïve\xspace}

\newcommand{\sr}{ghost staging\xspace}
\newcommand{\Sr}{Ghost staging\xspace}
\newcommand{\SR}{Ghost Staging\xspace}

\newcommand{\dgp}{direction-guided selection\xspace}
\newcommand{\Dgp}{Direction-guided selection\xspace}
\newcommand{\DGP}{Direction-Guided Selection\xspace}

\newcommand{\cp}{pipelining-based path extension\xspace}
\newcommand{\Cp}{Pipelining-based path extension\xspace}
\newcommand{\CP}{Pipelining-based Path Extension\xspace}

\usepackage{tikz}
\newcommand*\circled[1]{\tikz[baseline=(char.base)]{
            \node[shape=circle,draw,inner sep=0.4pt, fill=white, text=black] (char) {#1};}}

\newcommand{\shep}[1]{{\color{olivegreen}#1}}

\newcommand{\shephl}[1]{{\sethlcolor{pink}\hl{#1}}}

\newcommand{\shepref}[1]{{\color{magenta}{#1}}}
\newcommand{\shepindex}[1]{\label{shep:#1}\marginnote{{\color{magenta}\textbf{#1}}}}

  \usepackage[available]{usenixbadges}

\begin{document}


\newboolean{sheperding}
\setboolean{sheperding}{false}

\ifthenelse{\boolean{sheperding}}{
    
    \setcounter{page}{1}
    \setcounter{section}{0}
    \setcounter{figure}{0}
    
    \newrefsection 
}{
\iftrue 
\renewcommand{\shep}[1]{#1}
\renewcommand{\shephl}[1]{#1}
\renewcommand{\fcolorbox}[3]{#3}
\def\bigbox{\begingroup}
\def\endbigbox{\endgroup}
\renewcommand{\shepindex}[1]{}
\renewcommand{\shepref}[1]{}
\else
\fi 
}

\date{}

\title{\Large \bf \thiswork: A High-Throughput Multi-GPU System for \\ Graph-Based Approximate Nearest Neighbor Search}

\author{
  \textnormal{Sukjin Kim} \hspace{1em}
  \textnormal{Seongyeon Park} \hspace{1em}
  \textnormal{Si Ung Noh} \\
  \textnormal{Junguk Hong} \hspace{1em}
  \textnormal{Taehee Kwon} \hspace{1em}
  \textnormal{Hunseong Lim} \hspace{1em}
  \textnormal{Jinho Lee} \\
  \\
  Seoul National University
}

\maketitle

\begin{abstract}

Graph-based Approximate Nearest Neighbor Search (ANNS) is widely adopted in numerous applications, such as recommendation systems, natural language processing, and computer vision. 
While recent works on GPU-based acceleration have significantly advanced ANNS performance, the ever-growing scale of datasets now demands efficient multi-GPU solutions.
However, the design of existing works overlooks multi-GPU scalability, resulting in \naive approaches that treat additional GPUs as a means to extend memory capacity for large datasets.
This inefficiency arises from partitioning the dataset and independently searching for data points similar to the queries in each GPU.  
We therefore propose \thiswork, a novel multi-GPU framework designed to scale and accelerate ANNS for large datasets. 
First, we propose \cp, a GPU-aware pipelining mechanism that reduces prior work's redundant search iterations by leveraging GPU-to-GPU communication.
Second, we design \sr that leverages a representative dataset to identify optimal query starting points, reducing the search space for challenging queries. 
Finally, we introduce \dgp, a data selection technique that filters irrelevant points early in the search process, minimizing unnecessary memory accesses and distance computations.
Comprehensive evaluations across diverse datasets demonstrate that \thiswork achieves \textbf{$3.24\times$} geomean speedup and up to \textbf{$5.30\times$} speedup on 95\% recall rate over state-of-the-art multi-GPU-based ANNS frameworks.

\end{abstract}

\section{Introduction}
Various fields, such as computer vision~\cite{mining-visual-phrases, lifelong-visual-maps, clustering-imgs}, recommendation systems~\cite{anns-recommender}, natural language processing~\cite{gtm, why-nnlm-work}, and information retrieval~\cite{lsa}, 
utilize datasets consisting of multi-dimensional vectors representing larger data entities (e.g., image, text).
These representations are used to efficiently retrieve data relevant to the input queries in various applications. 
One solution to finding $k$-closest data points (vectors) is $k$-Nearest Neighbor Search, where the $k$-closest data points are selected from the entire dataset based on the user-defined similarity metric, commonly the L2 distance. 
However, as the dataset grows, finding the exact solution becomes increasingly difficult due to the curse of dimensionality~\cite{ann-curse-dim, approx-closest-queries, anns-highdim}.
To address this challenge, Approximate Nearest Neighbor Search (ANNS) has been widely adopted, especially by products such as vector databases~\cite{pinecone, mongodb, weaviate, faiss, chroma, milvus}.
ANNS performs nearest neighbor search on large datasets with reasonable execution times while maintaining high accuracy.

Among various ANNS solutions~\cite{pq, opq, opt-trees, qals-hashing}, graph-based methods~\cite{nsw, hnsw, efanna, nsg, hcnng} have gained significant attention due to their ability to effectively represent neighbor relationships between data points. 
This representation allows for evaluating fewer data points while achieving higher accuracy compared to alternative methods~\cite{wang2021comprehensive}. 
Consequently, there has been a growing body of work on optimizing graph-based ANNS performance across diverse hardware platforms.
While CPU-based optimizations~\cite{grasp, graphreorder_cache, finger, quickadc, adsampling, pecann, parlayann, highdim} offer algorithms that maintain accuracy, they are limited by the number of available cores, hindering their scalability when processing large, high-dimensional datasets that demand extensive computation and memory access.

To address such a computational burden, recent works~\cite{song, ggnn, cagra, bang, billscalegpu, nsw, diskann} propose GPU-optimized search techniques to leverage the high-performance capabilities of GPUs. 
A prominent example is CAGRA~\cite{cagra}, which accelerates single-GPU search by efficiently computing the L2 distance between queries and data points. 
It further leverages hash tables to eliminate redundant computations while optimizing for the GPU's thread and memory hierarchy. 
Another example is GGNN~\cite{ggnn}, which partitions large datasets into smaller, independent graphs that fit within a single GPU.
This enables data-parallel searches across multiple GPUs.

While these approaches demonstrate significant performance improvements over CPU-based solutions, we identify key bottlenecks that hinder their ability to fully exploit the potential of GPUs:
\vspace{1.2em}
\begin{itemize}
    \item \textbf{Existing multi-GPU solutions suffer from low scale efficiency.} 
    Prior works~\cite{ggnn, billscalegpu} divide the dataset across GPUs to support ANNS on large datasets with GPUs, with each GPU independently processing queries (i.e., graph sharding). 
    While this approach enables handling large datasets on GPUs, it requires each query to be processed multiple times across different GPUs, resulting in a low scale efficiency.
   
    
    \item \textbf{Majority of random initial nodes turn out to be unnecessary later.} 
    To quickly search for near-optimal data points, existing approaches~\cite{cagra, nsw, diskann} rely on starting searches from numerous random initial points. 
    However, due to the underlying beam search method, most of the neighbors explored from those random initial points are quickly discarded within a few iterations, leaving only descendants of the few best points. 
    This incurs too much computational and memory overhead for the search, which is unnecessary for the final result.
    
    \item \textbf{Each iteration mandates too much overhead.}
    Visiting a vertex in a proximity graph yields distance computation with all its neighbors. 
    Given that the typical degree of proximity graphs is around a few tens, this is a huge burden to the search.  
    However, many of these neighboring data points are not selected as part of the top-$k$ results, leading to unnecessary memory access and computation that hinders overall efficiency.
\end{itemize}


Based on these observations, we introduce \textbf{\thiswork}, a multi-GPU framework designed to support graph-based ANNS on large datasets with scalable performance with minimal accuracy loss. 
\thiswork is carefully designed to address the limitations above while exploiting the parallelism of GPU resources.

To address the scalability limitations of sharding-based multi-GPU search, we leverage our finding that each shard's seemingly independent search operations can be optimized by sharing intermediate search results with the following shard.
We propose \textbf{\cp}, a mechanism that passes the search results of each shard to the next GPU in a pipelined manner.
This enables subsequent GPUs to start their search from data points closer to the query within their shard, effectively reducing the number of search iterations only with negligible inter-GPU communication cost.
One limitation of \cp lies in the first stage, whose search still has to be done from scratch.
Our second scheme \textbf{\sr} addresses this issue by selecting more optimal starting points for each shard by prioritizing data points closer to the query, thereby improving the efficiency of the first stage.
Finally, we propose \textbf{\dgp}, which skips distance calculations for neighbors whose direction from the parent node significantly deviates from the direction toward the query. 
This approach further minimizes unnecessary distance computations, reducing both computational and memory overhead.

\shep{
\shepindex{S3}
We implement the proposed techniques in \textbf{\thiswork} and conduct extensive experiments on large datasets. \thiswork improves the performance by $3.24\times$ geomean speedup compared to the state-of-the-art GPU-based ANNS baselines at $95\%$ recall rate, demonstrating its scalability and effectiveness. 
}

\section{Background}

\subsection{Approximate Nearest Neighbor Search}

Given a dataset $\mathcal{D} = \{x_0, x_1, \cdots, x_{n-1} \mid x_i \in \mathbb{R}^d \}$ with $n$ data points represented as $d$-dimensional vectors, and a query $q \in \mathbb{R}^d$, $k$-Nearest Neighbor Search ($k$-NNS) identifies $k$ data points $N_k(q)$ that satisfies the following:
\begin{equation}
    N_k(q) = \operatorname*{argmin}_{\mathcal{S} \subset \mathcal{D}, |\mathcal{S}| = k} \sum_{x_i \in \mathcal{S}} dist(q, x_i),
\end{equation}
where $dist(q, x_i)$ is the similarity measure between the query and data point, typically the L2 distance. 
As the size of $\mathcal{D}$ grows, performing exact $k$-NNS becomes computationally expensive due to the substantial memory access and computation overhead. 
To address this, many applications adopt approximate nearest neighbor search (ANNS), which employs efficient algorithms to approximate $k$-NNS results with significant reductions in execution time~\cite{improving-anns}.

Among the various ANNS algorithms, graph-based ANNS~\cite{nsw, hnsw, efanna, nsg, hcnng} has been widely used due to its ability to effectively capture the relationships (distances) between data points~\cite{wang2021comprehensive}. 
Graph-based methods construct a proximity graph $\mathcal{G}$ where nodes represent data points and edges encode distances between them, allowing for compact representations of relationships and facilitating efficient graph traversal to approximate $k$-NNS results.
Assuming that $\mathcal{G}$ is pre-constructed, the proposed \thiswork targets accelerating the graph search process itself, addressing key inefficiencies and scaling challenges associated with large-scale datasets.

\subsection{Approximate Nearest Neighbor Search Method on GPUs}
\label{sec:back}

\begin{figure}[t]
    \centering
    \includegraphics[width=.98\columnwidth]{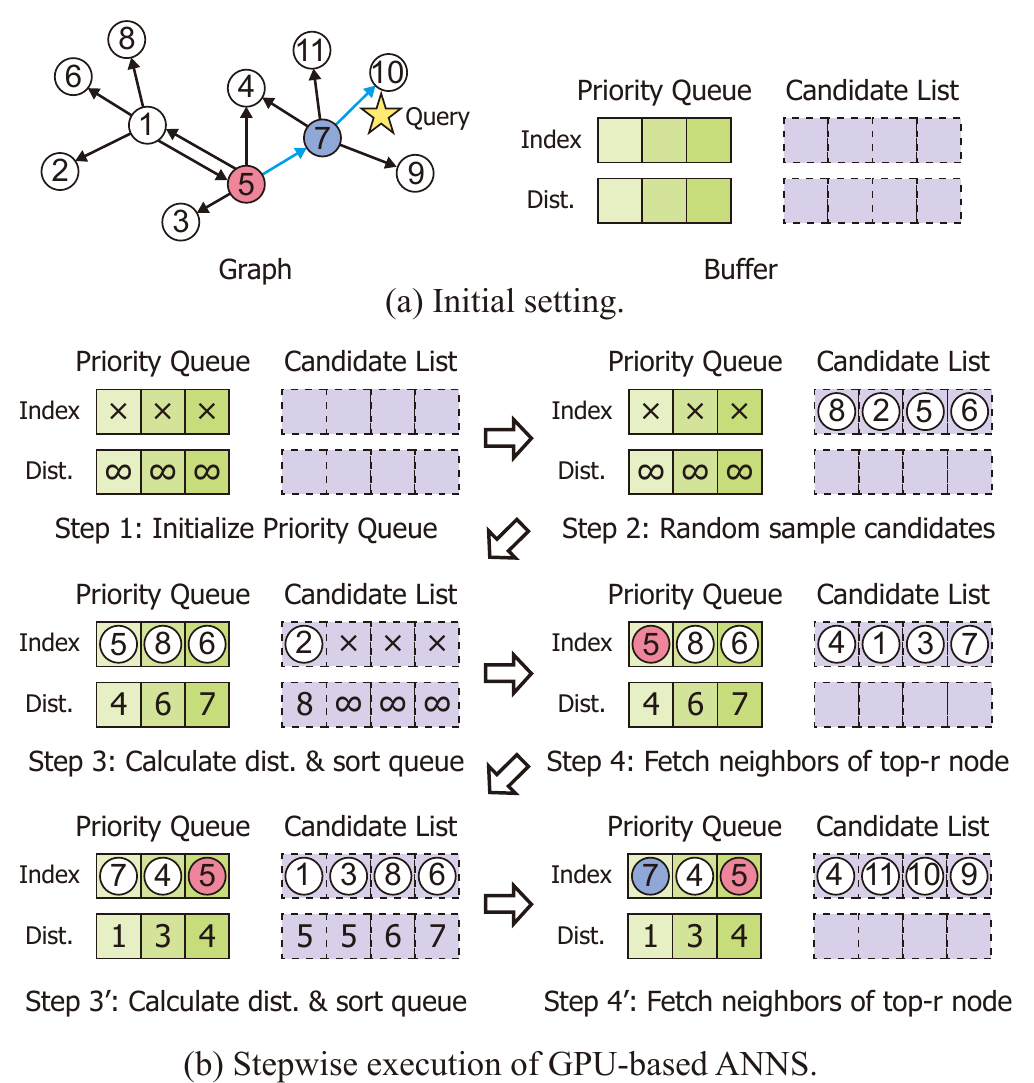}
    \caption{Overview of approximate nearest neighbor search on GPUs.
    }
    \label{fig:bg}
\end{figure}

Graph-based Approximate Nearest Neighbor Search (ANNS) algorithms have gained significant traction due to their ability to efficiently navigate complex datasets by leveraging graph structures. 
We briefly explain the overall search algorithm~\cite{cagra} on a constructed graph $\mathcal{G}$, as illustrated in \cref{fig:bg}a, to provide context for our optimizations. 
Note that many graph-based search algorithms~\cite{ggnn, song, bang} follow a similar workflow. 

The graph-based search operates on two key data structures as shown in \cref{fig:bg}a: 
\begin{itemize}
    \item A \textit{priority queue} $p = \{p_0, p_1, \dots, p_{l-1}\}$ of size $l$, which stores the top-$l$ ($k \leq l$) intermediate nodes sorted by their distances to the query $q$.
    \item A \textit{candidate list} $c = \{c_0, c_1, \dots, c_{m-1}\}$ of size $m$, which acts as a buffer to hold the neighbors of the nodes being processed.
\end{itemize}

Given these structures, the algorithm iteratively refines the search process as depicted in \cref{fig:bg}b:
\begin{enumerate}[label=\textbf{Step \arabic*}, leftmargin=*]
    \item The algorithm begins by initializing the priority queue $p$ with dummy indices with $\infty$ distances.
    \item The candidate list $c$ is populated with $m$ randomly selected nodes from the graph. 
    \item The distance $dist(q, c_i)$ is computed for each candidate node $c_i \in c$ using:
        \begin{equation}
        dist(q, c_i) = \| q - c_i \|_2,
        \end{equation}
        where elements of the priority queue $p$ and the candidate list $c$ are sorted together based on the distances, ensuring that the closest nodes appear at the top of $p$.
    \item The neighbors of the top-$r$ ($r \leq l$) nodes in the priority queue are fetched from $\mathcal{G}$ and the candidate list $c$ is populated with these nodes.
    \end{enumerate}
    Steps 3 and 4 are repeated until the priority queue receives no new entries or the pre-defined maximum number of iterations is reached. Then, the top-$k$ nodes in $p$ are returned as the approximate nearest neighbors. 


\begin{figure}[t]
    \centering\shepindex{S1}
    \fcolorbox{olivegreen}{white}{
        \includegraphics[width=.97\columnwidth]{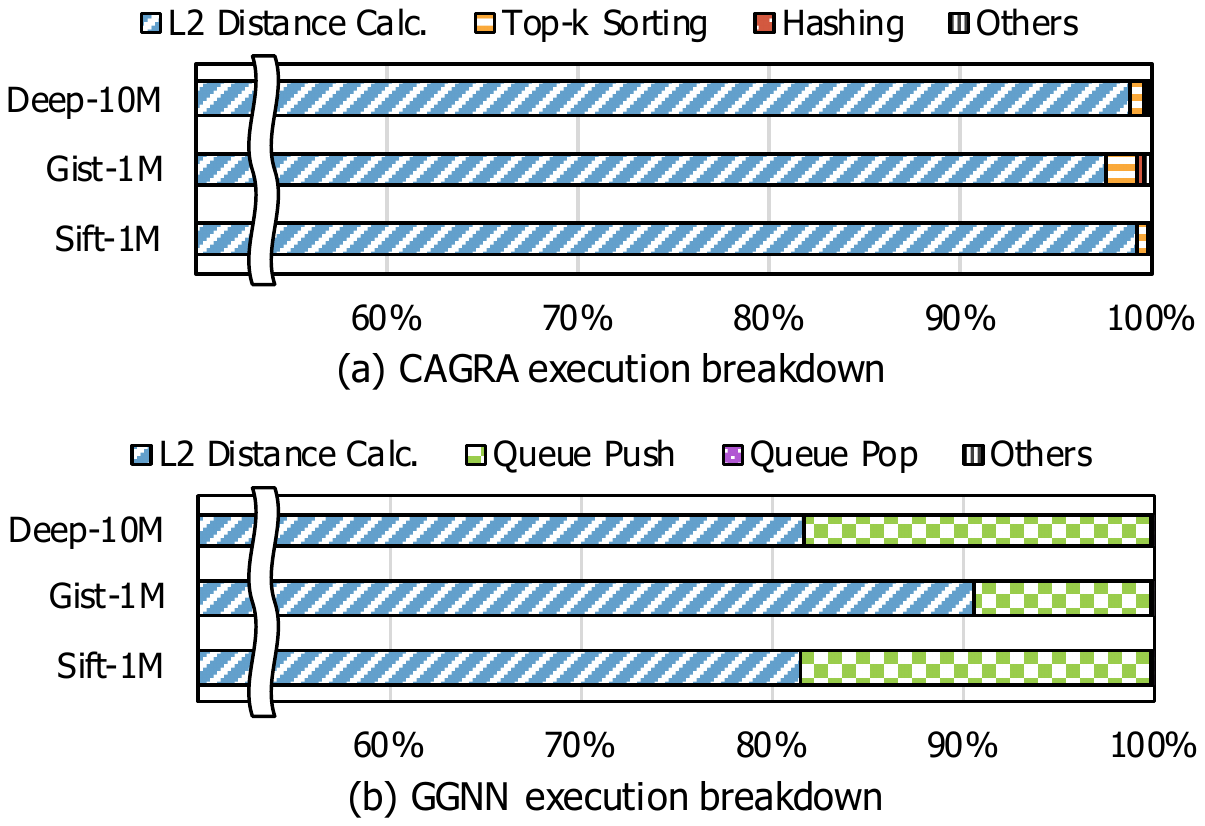}
    }
    \caption{Execution time breakdown analysis of baseline ANNS. 
    }
    \label{fig:moti}
\end{figure}

This algorithm is based on carefully built proximity graphs, which encourage reachability (all vertices are reachable starting from any vertex) and convexity (avoids falling into local minima)~\cite{ggnn, cagra}.
Based on such characteristics, the search algorithm aims for faster convergence towards the global minima by maintaining the priority queue that stores data points with small L2 distances and quickly dropping long-distance points.

\shep{
\shepindex{S1}
Because the vector dimension typically ranges from a few tens to several hundreds, L2 distance calculation dominates the search time due to the overhead of loading and processing high-dimensional vectors.
According to our measurements depicted in \cref{fig:moti}a, when using the current state-of-the-art method CAGRA~\cite{cagra}, L2 distance calculation accounts for over 95\% of the total search time, regardless of the dataset.
To further analyze this trend across different methods, we also performed a breakdown of GGNN~\cite{ggnn} and depicted in \cref{fig:moti}b found that over 80\% of the time was still spent on distance calculations.
}
Therefore, the key to high-throughput ANNS is to reduce the number of L2 distance calculations.
For a search that converged in $i$ iterations on a $j$-proximity graph (i.e., each vertex has $j$ neighbors), the rough count of L2 distance calculation is $i\times j \times r$ (including duplicates).
To achieve high throughput, \thiswork mainly aims to reduce the number of search iterations $i$ and number of examined neighbors out of $j$, while not sacrificing the search accuracy. 


\section{\thiswork Design}
\label{sec:design}


\subsection{\CP}

\subsubsection{Scalability of Performance on Multiple GPUs}
\label{sec:cp_moti}

\begin{figure}[t]
    \centering\shepindex{S1}
    \fcolorbox{olivegreen}{white}{
        \includegraphics[width=.97\columnwidth]{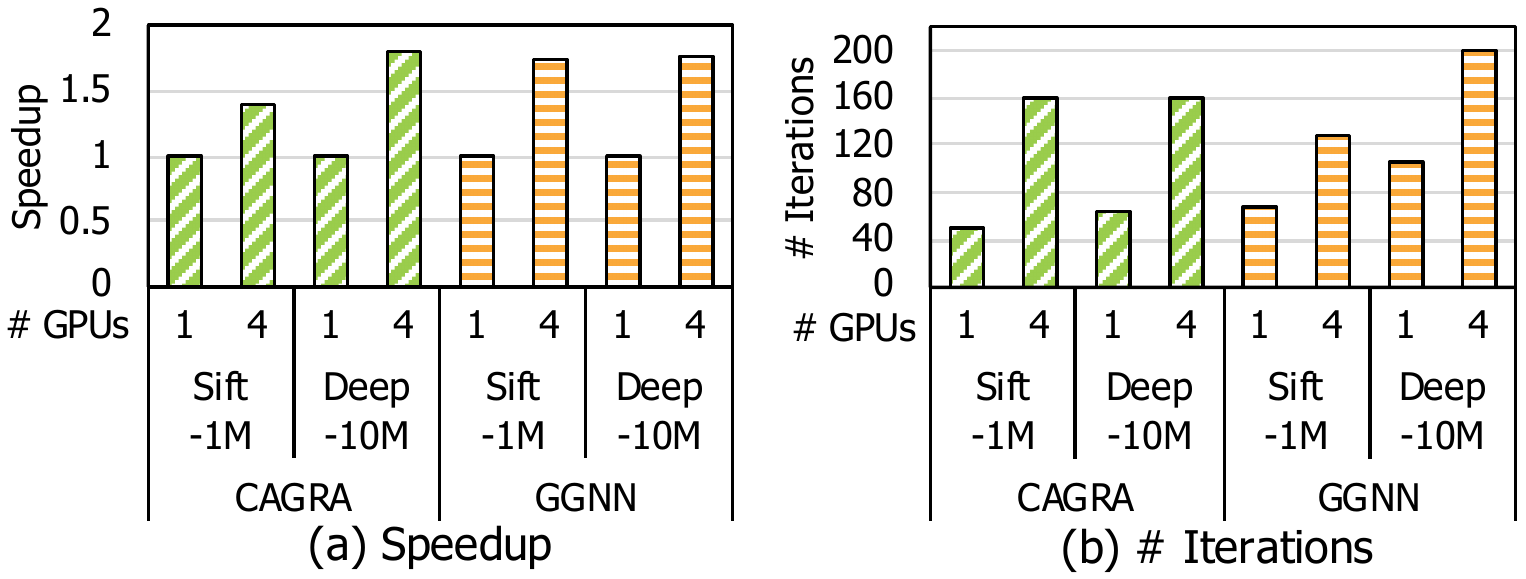}
    }
    \caption{Performance scalability of prior work on Sift-1M and Deep-10M datasets with $4$ GPUs.
    }
    \label{fig:cp_moti}
\end{figure}


\shep{
\shepindex{S1}
To diagnose the existing multi-GPU search method based on sharding~\cite{ggnn}, we compare the total number of search iterations between single- and multi-GPU setups using a multi-GPU extension of CAGRA and GGNN, as shown in \cref{fig:cp_moti}a.
The results indicate that the sharding-based approach scales inefficiently across multiple GPUs in both methods.
For the Sift-1M dataset, using 4 GPUs yields only a 1.39$\times$ speedup in CAGRA, resulting in a scaling efficiency of approximately 35\%.
GGNN shows similar trends, achieving about a 1.7$\times$ speedup on both datasets, which corresponds to a scaling efficiency of around 43\%.
}
This inefficiency arises from the fact that the multi-GPU baseline performs a separate local search in each GPU's shard for every query. 
While this simple query-parallel design provides an easy parallelized solution for large datasets, 
the number of iterations needed does not reduce linearly with the size of a shard. 
\shep{
\marginpar{\shepindex{S1}}
As a result, the number of total iterations to complete a search increases with more number of shards, as depicted in \cref{fig:cp_moti}b.
In the case of the Sift-1M dataset in CAGRA method, the total iteration over all the shard scales with the number of shards results in 4$\times$ total iteration.
In GGNN, increase of total iterations is slightly better at around 2$\times$, but not enough to compensate for the overall scalability loss.
This highlights the need for an alternative solution to achieve efficient performance. 
}

\subsubsection{\CP Design}

\begin{figure}[t]
    \centering\shepindex{S6}
    \fcolorbox{olivegreen}{white}{
      \includegraphics[width=.98\columnwidth]{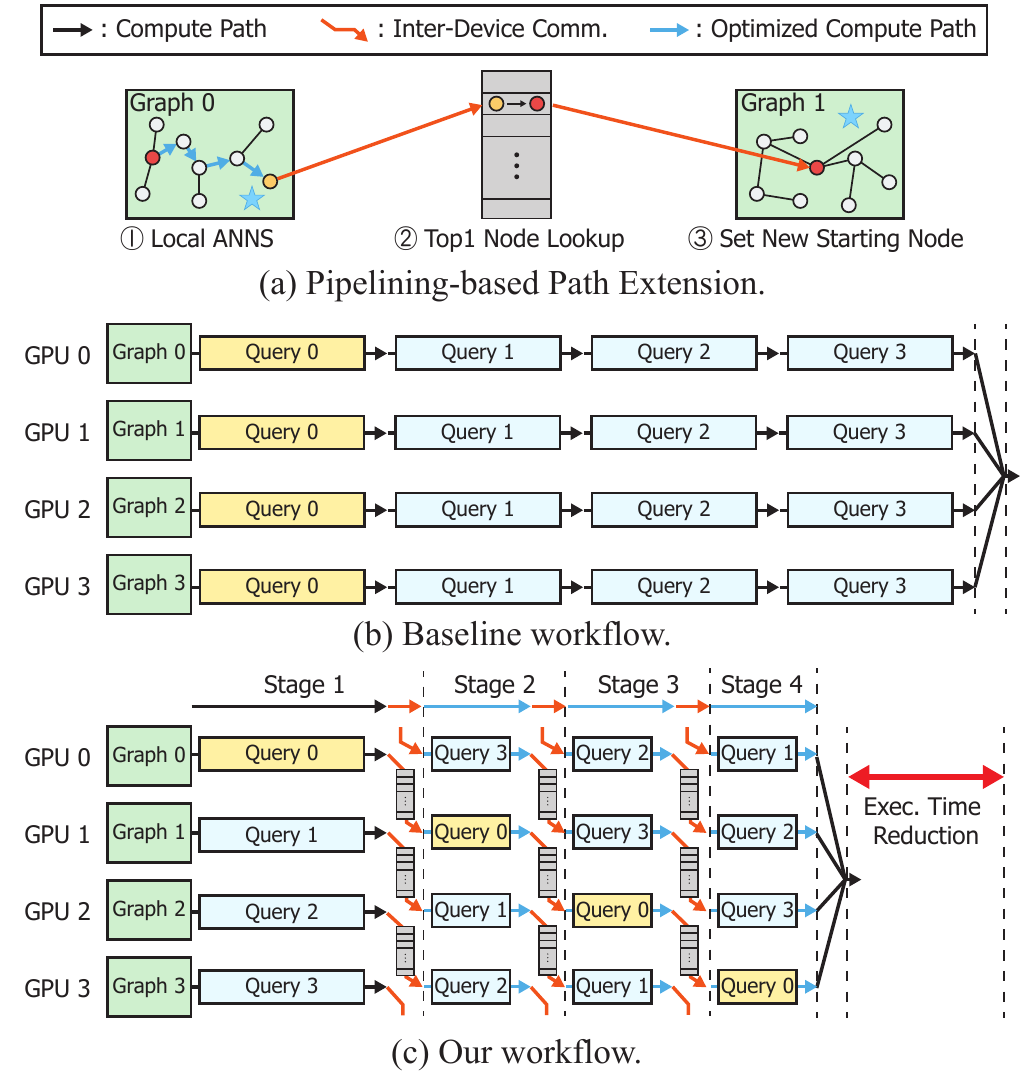}
    }
    \caption{Illustration of the \Cp design.
    }
    \label{fig:cp}
\end{figure}
To amend the issue identified above, we propose \emph{\cp}, a multi-GPU solution that enhances scalability by leveraging intermediate search results from other GPUs in a pipelined manner. 
As illustrated in \cref{fig:cp}a, the key idea is to create inter-shard edges from each node to the closest node in the adjacent shard.  

The baseline sharding~\cite{ggnn} (\cref{fig:cp}b) begins by randomly partitioning the input dataset into independent shards, with the number of shards matching the number of GPUs ($4$ shards in this example). 
After allocating these shards to each GPU, independent graphs are built for each shard/GPU. 
\shep{
\marginpar{\shepindex{S2}}
Note that in this baseline approach, there is no edge between the nodes in different graphs.
} 
Once the graphs are built, the baseline performs independent ANNS for all queries by starting from randomly selected nodes within each graph. 
Upon completing the search for all queries, each GPU stores the top-$k$ data points for each query specific to its graph. 
For instance, after the searches are finished, $N \times k$ candidate data points are outputted across $N$ GPUs. 
These results are then offloaded to the CPU for reduction to get the final top-$k$ data points for each query. 
While this method provides an easily parallelizable solution with almost no communication, it requires more search iterations as revealed in \cref{sec:cp_moti} and thus scales inefficiently. 

\shep{
\shepindex{S2}
In the proposed \cp design illustrated in \cref{fig:cp}c, the dataset is randomly partitioned in the same way as the baseline sharding approach to build independent graphs $G_i$ for the shard on GPU $i$. 
}
However, after the graph construction, \cp takes a step further by creating uni-directional node-to-node connections between the shards of adjacent GPUs as shown in \cref{fig:cp}a. 
We define $G_i$ to be adjacent to shards graph $G_{(i+1)\% N}$ for $N$ GPUs, which form a ring topology.
From each vertex in a shard, the nearest neighbor vertex is found in the adjacent shard, and an inter-shard edge is added between them.
\shep{
\shepindex{S2}
For example, the $0$th GPU's graph connects all of its nodes to the closest node in the $1$st GPU's graph, and nodes in the $3$rd GPU to the $0$th GPU, forming a ring-like uni-directional connection among graphs/shards. 
}
These inter-shard edges can be expressed as a mapping $I$:
%
%
\begin{align}
\quad I(u) = \operatorname*{argmin}_{w \in G_{(i+1) \% N}} dist(u, w)\qquad  \forall u \in G_i. 
\end{align}
These connections are constructed once during the graph build phase and can be reused across searches. 
As detailed in \cref{sec:eval:build}, the additional overhead for creating these inter-shard edges is minimal.

 To perform a search with the interconnected graph, each GPU is assigned a $|Q|/N$ chunk from the set of queries $Q$ and starts the search as in the baseline sharding.
After each query has converged in all GPUs, a part of the final local result $Z$ is taken. 
From each of its vertices $z\in Z$, it continues searching from $I(z)$, which naturally takes place in the adjacent shard as if the search path has been extended.
Because $I(z)$ has been designed to be close to the query, the search converges significantly faster than the baseline in fewer iterations.
\shep{
\marginpar{\shepindex{S6}}
For example, in \cref{fig:cp}c, when searching with query batch 0 (yellow box), the search process begins on GPU 0 with graph 0.
After the search completes on GPU 0, the result is passed to GPU 1, which continues the search on graph 1.
As shown in the figure, the search time in stage 2 (blue arrow) is reduced compared to that of stage 1 (black arrow).
This process is executed simultaneously on all GPUs and is repeated in a pipelined manner until every GPU has processed the queries originating from all other GPUs.
}
Then the results are reduced in the CPU, similarly to the baseline.

\shep{
\shepindex{S2}
As discussed in \cref{sec:disc:pp_cost}, because the only data transfer involved is the local result between the adjacent GPUs, the communication overhead is almost negligible compared to the substantial reduction in search time.
}

%

\subsection{\SR}

\subsubsection{Execution Time Breakdown after \CP}

\begin{figure}[t]
    \centering
    \includegraphics[width=.96\columnwidth]{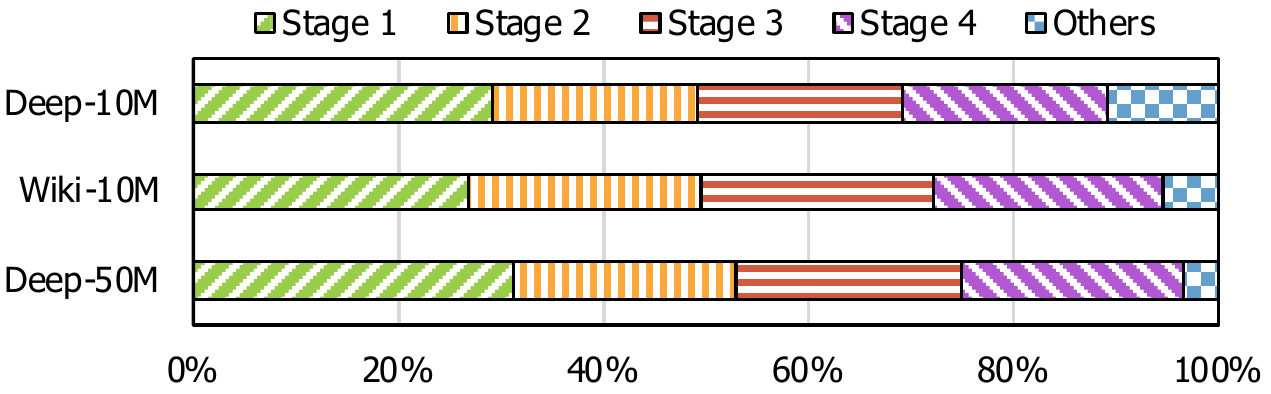}
    \caption{Execution time breakdown analysis with 4 GPUs after applying the \cp design according to the pipeline stages. 
    }
    \label{fig:sr_moti}
\end{figure}

\cref{fig:sr_moti} illustrates the execution time breakdown of searching with \cp. 
As expected, the first search stage of \cp is the primary bottleneck.
For example, the first search stage in the Deep-50M dataset takes up to 31\%, while the other stages consume only up to 22\%.

This bottleneck arises as the first search stage does not benefit from \cp and starts from random, potentially distant data points. 
As a result, more iterations are required to converge on the appropriate top-$k$ candidates. 
As all GPUs are performing the initial stage of their own query, reducing this overhead could bring additional speedup.

\subsubsection{\SR Design}

\begin{figure}[t]
    \centering
    \includegraphics[width=.98\columnwidth]{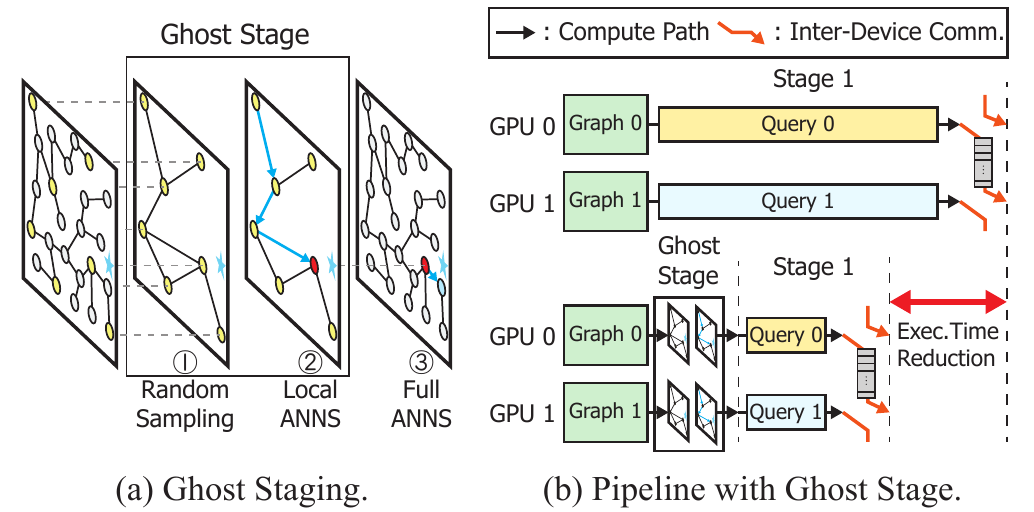}
    \caption{Illustration of the \sr design.
    }
    \label{fig:sr}
\end{figure}

\shep{
\shepindex{S1}
Motivated by \cp's ability to identify better starting points for subsequent GPU search stages, we propose \emph{\sr}, a kind of hierarchical search~\cite{hnsw} solution for improving the initial search stage on each GPU. 
}
\Sr adds a small auxiliary shard before the first stage, such that a high-quality data point close to the query can be located only in a few iterations.
This approach significantly reduces the number of iterations needed during the initial search stage.

\cref{fig:sr} illustrates how \sr reduces the number of search iterations during the first stage while maintaining accuracy. 
To create an auxiliary shard, \circled{1} \sr randomly samples a fixed number of random data points within a shard, referred to as ghost nodes. 
These ghost nodes are connected based on distances, creating a lightweight network.
In addition, the inter-shard edges are added between the ghost nodes and the original nodes.
Due to the relatively small number of ghost nodes, the preparation for \sr only adds negligible overhead to the graph build process (see \cref{sec:eval:build}).
During \sr, \circled{2} a fixed number of starting points are chosen among the ghost nodes and the search algorithm from \cref{sec:back} is applied to locate ghost nodes near the query within a few iterations. 
\circled{3} Once the relevant ghost nodes are identified, the search transitions to the original graph similar to \cp.

The high performance of \sr stems from ghost nodes acting as central hubs and their interconnections as highways within the expansive original graph. 
Moreover, \sr maintains high accuracy by locally searching the original graph for the remaining iterations. 
This is highlighted by comparing the search process of the baseline method and \sr in \cref{fig:sr}b.
To reach the same data point close to the query, the baseline has to traverse many vertices due to the original graph's size. 
On the other hand, during \sr, taking a single step forward among ghost nodes has a similar effect of passing through multiple original graph data points. 
Even if \sr overshoots and bypasses the query, it can backtrack by the remaining search on the original graph.

\subsection{\DGP}

\subsubsection{Analysis on Unused Distance Calculation}

\begin{table}[t]
\small
    \centering 
    \caption{Unused Distance Calculation}
    \vspace{0.5em}
    {
    \resizebox{.97\columnwidth}{!}
    {
    \setlength{\tabcolsep}{3pt}
    \shepindex{S6}
    \fcolorbox{olivegreen}{white}{
    \begin{tabular}{lccc}
        \toprule
     \textbf{Dataset} & \textbf{$\#$Total Visits} & \textbf{$\#$Discarded Visits} & \textbf{Ratio}\\  
    \midrule
     Sift-1M~\cite{datasets-anns, pq} & 2.32E+7 & 2.00E+7 & 86.2\% \\
     Gist-1M~\cite{datasets-anns, pq} & 3.17E+6 & 2.81E+6 & 88.9\% \\
     Deep-10M~\cite{deep} & 2.68E+7 & 2.28E+7 & 85.0\% \\
    \bottomrule
    \end{tabular}
    } 
    } 
    }
    \label{tab:dgp_moti}

\end{table}

Despite reducing the search iterations with \cp and \sr, the search procedure remains heavily dominated by the numerous L2 distance calculations.
This partially stems from having to add all neighbors of each top-$r$ node in the priority queue (\cref{sec:back}), where the number of neighbors is usually in the order of a few tens~\cite{cagra}.
While such a large number of neighbors is necessary to ensure reachability and convexity, it often introduces unnecessary computational overhead.
During the search, we found that the majority of the data points added to the priority queue never reached the top-$k$ region of the priority queue and were eventually discarded (unused) without being considered again because they were too far from the query compared to the other candidates. 

In \cref{tab:dgp_moti}, we quantitatively measured the portion of the considered nodes, 
\shep{
\marginpar{\shepindex{S6}}
where \textbf{\#Total Visits} denotes the total number of nodes in the proximity graph that are accessed during the search,
and \textbf{\#Discarded Visits} represents the nodes that are visited but never remain in the candidate buffer until the end of the search process to be included in the final top-$k$ results.
As depicted in the table, we find that the ratio of discarded node visits, which lead to unnecessary distance calculations, can exceed 80\%.
}
This indicates an interesting opportunity for additional optimization, where we could remove a substantial portion of the unused calculations if we can identify them in advance.

\subsubsection{\DGP Design}

\begin{figure}[t]
    \centering
    \includegraphics[width=.98\columnwidth]{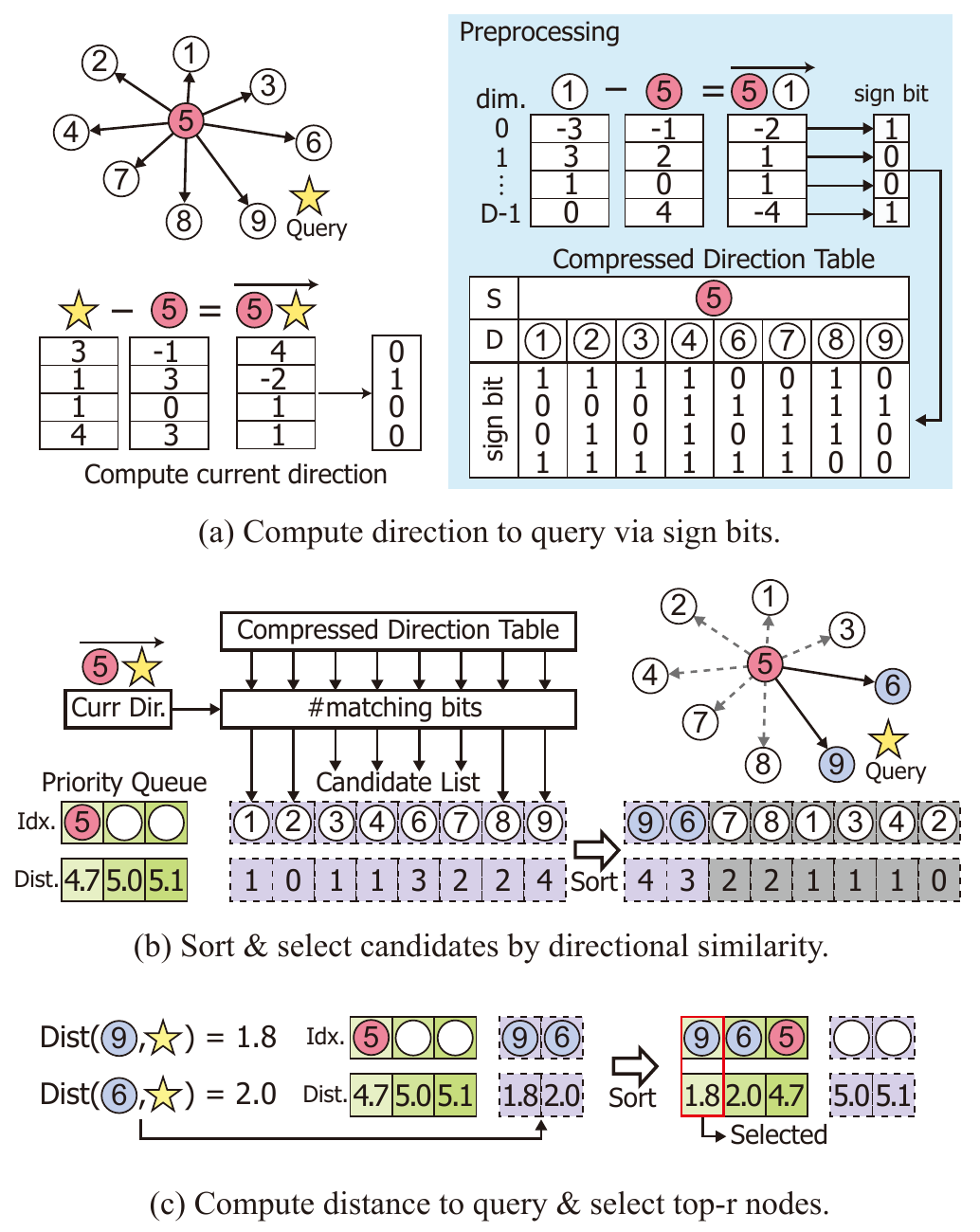
    }
    \caption{Illustration of the \dgp design.
    }
    \label{fig:dgp}
\end{figure}

We propose \emph{\dgp}, an optimization technique designed to accelerate the search process within each iteration. 
\Dgp allows the search algorithm to bypass data points that are significantly unaligned with the query's direction relative to the current target node, as shown in the top graph in \cref{fig:dgp}. 
By leveraging a set of lightweight operations including direction table reads and local sorting, \dgp significantly reduces expensive vector reads and distance calculations.

\shep{
\shepindex{S5}
The \dgp process operates as follows:
(a) During the off-line preprocessing phase, a compressed direction table is generated, storing the sign bits of the differences between the source node and its neighbors in the proximity graph.
Each sign bit vector encodes the rough direction of an edge~\cite{pouransari2020least} and serves as a fast proxy to guide traversal during run-time search on the GPU.
}
(b) On each iteration of the search, the sign bits of the direction between the query and the visiting node (node $5$ in the example) are calculated. 
Then the number of matching bits is counted for each neighbor using the compressed direction table.
The neighbors are sorted by their number of matching bits in descending order, and the top-$n$ candidates are selected.
In the example, with $n = 2$, nodes $9$ and $6$ are selected.
(c) The distance is calculated between the query and the selected neighbors (node $9$ and $6$ in the example), and the priority queue elements are sorted by the order of their distances (e.g., nodes $9, 6, 5$).
Finally, top-$r$ nodes from the priority queue are selected (e.g., node $9$ with $r=1$) and they become the next visiting nodes.
To maintain accuracy, the last few \MAPAE{50\% = few?}iterations are regarded as cool-down phase, and the search is conducted without the selection.
We use 30\% of the max iteration as the default value.

\Dgp effectively reduces the number of distance calculations while maintaining high accuracy. 
This is achieved by prioritizing candidates most aligned with the query’s direction, significantly narrowing the search space.
There is a small chance that \dgp can drop meaningful candidates, because of the compression error in the direction table.
However, due to the reachability and convexity of the proximity graphs, this would typically result in an increased number of iterations until convergence instead of accuracy.
We empirically found that the accuracy drop is almost negligible while the gain from the reduction in candidates dwarfs the increase in the number of iterations. 

\section{Implementation Details}
\thiswork's search kernel has been implemented based on the CAGRA search kernel, adopting its query-per-thread-block approach along with GPU-friendly bitonic sorting and hash tables.
To maximize the benefits of fast register communication within a warp, we set the threadblock size to be equal to that of a warp (i.e., 32 threads).

To minimize the communication overhead in \cp, the total set of queries is sent to all devices, which has a negligible size compared to the dataset.
Each device processes independent graphs generated from its respective shard and maintains a unique inter-shard edge table for its linked shard. 
At runtime, only a partial chunk of the query batch is processed, and the result is forwarded to the next GPU device. 

While the number of results sent per query is a tunable parameter, we empirically choose to send only one to minimize communication overhead.
Additionally, for \cp, a shard-to-shard lookup table needs to be precomputed.
Every node within a single shard acts as a query to perform a search in the adjacent shard.
The top-1 result from the search is stored in the lookup table for \cp.

\Sr can be implemented similarly to \cp, except that instead of connecting neighboring shards, it connects the extracted smaller auxiliary shard with the original shard.

In \dgp, during processing of the sign bit table, all 1-bit signs are packed into one uint32. 
To efficiently compress the vector between the current node and the query, each thread subtracts the vector's elements and converts them into 1-bit values.
During the runtime search process, the \texttt{\_\_shfl\_xor\_sync()} intrinsic is used within a warp to enable efficient processing by performing fast intra-warp shuffle operations.  
Next, the precomputed sign bit table is looked up to retrieve the approximate sign bits of neighboring vectors. 
The similarity with the query vector is then computed by performing bit-wise XOR operations on uint32 units, followed by a \texttt{\_\_popcll()} intrinsic call to count the total number of different bits. 
Finally, a min-sort is performed to find the node with the highest similarity.
As mentioned above, \dgp requires a precomputed compressed sign bit table generated on the CPU using multi-threading, where each thread handles the edges of a single parent node.
For each neighbor, element-wise comparisons are performed, and the results are compressed into a uint32 using bit-wise shift and OR operations.

\section{Evaluation}

 \subsection{Experiment Setup}
\begin{table}[t]
    \centering 
    \caption{Datasets used in evaluation}
    \vspace{0.5em}
    {
    {
    \setlength{\tabcolsep}{3pt}
    \begin{tabular}{llccc}
        \toprule
     \textbf{Target} & \textbf{Dataset} & \textbf{Dim. ($d$)} & \textbf{Size ($n$)} & \textbf{Type} \\
    \midrule
    \multirow{3}{*}{Single GPU} 
        & Sift-1M~\cite{datasets-anns, pq} & 128 & 1M & float \\
        & Gist-1M~\cite{datasets-anns, pq} & 960 & 1M & float \\
        & Deep-10M~\cite{deep} & 96 & 10M & float \\
    \midrule
    \multirow{3}{*}{Multi-GPU} 
        & Wiki-10M~\cite{wiki} & 768 & 10M & float \\
        & Deep-10M~\cite{deep} & 96 & 10M & float \\
        & Deep-50M~\cite{deep} & 96 & 50M & float \\
    \bottomrule
    \end{tabular}
    } 
    }
    \label{tab:dataset}
\end{table}

\textbf{Environment.}
\label{sec:eval:env}
We have implemented and evaluated \thiswork on a server with four NVIDIA RTX A6000 GPUs and an AMD EPYC 9124 16-Core CPU, where two GPUs are connected via NVLink Bridge~\cite{nvbridge}, and each GPU pair is linked through a host PCIe switch.
All environments run on Ubuntu 22.04 with CUDA version 12.1 and PyTorch 2.4.1.

\noindent\textbf{Baselines.}
The following CPU and GPU baselines were chosen for evaluation. 
\begin{itemize}
    \item \textbf{CAGRA}~\cite{cagra} is the state-of-the-art GPU framework for graph-based ANNS. CAGRA proposes a heuristically optimized proximity graph for parallel search operations and incorporates techniques such as warp splitting and forgettable hashing to enhance search performance. We used the official implementation of the authors, where we extended it for multi-GPU settings (denoted as `CAGRA w$\slash$ Sharding'). 
    \item \textbf{HNSW}~\cite{hnsw} is a popular graph-based ANNS implemented for CPUs. HNSW introduces a hierarchical proximity graph, with each layer structured as an NSW~\cite{nsw} graph representing a subset of points. The search operation begins at the top layer and traverses through the hierarchy, achieving improved performance and accuracy. Because HNSW only supports execution on CPUs, we evaluated HNSW only in the single-GPU environment for fair comparison. We utilized 64 threads on CPUs for evaluation.
    \item \textbf{GGNN}~\cite{ggnn} is another GPU implementation for graph-based ANNS. GGNN builds on the HNSW-inspired proximity graph, specifically optimizing it to leverage massive parallelism during the graph construction phase. Additionally, GGNN enhances search performance by efficiently utilizing shared memory and enabling parallel operations on data structures. To the best of our knowledge, GGNN is the only graph-based ANNS framework that supports multiple GPUs out of the box, where it uses the sharding method.
   
\end{itemize}

\begin{figure*}[t]
    \centering
    \includegraphics[width=.97\textwidth]{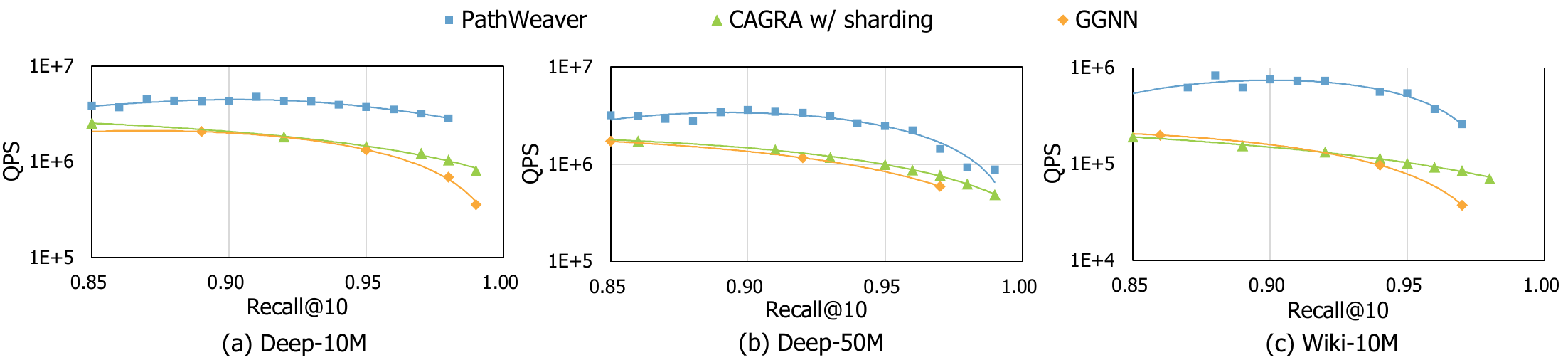}
    \caption{Performance comparison on multi-GPU environment. 
    }
    \label{fig:eval:multi}
\end{figure*}

\begin{figure}
    \centering
    \includegraphics[width=.99\columnwidth]{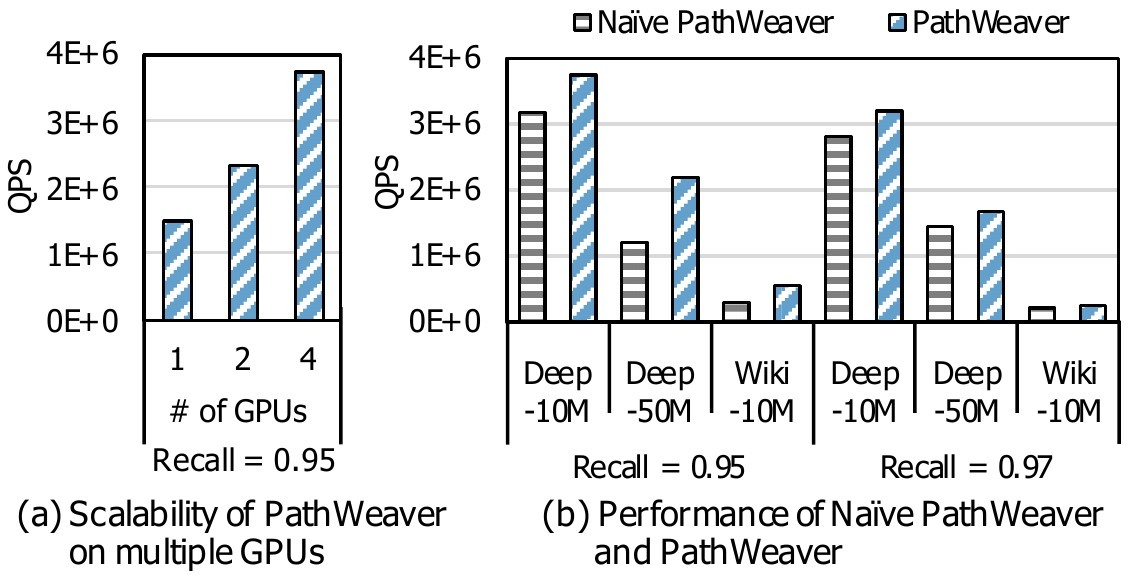}
    \caption{Performance comparison of Recall@10 between the \naive and pipeline method on multiple GPUs.}
    \label{fig:eval:multi_scale}
\end{figure}

\textbf{Datasets.}
We evaluate our method on a total of six datasets in which the vector dimension ranged from 96 to 960 as reported in \cref{tab:dataset}.
We analyzed single-GPU performance using datasets of varying sizes (1M, 10M) and dimensions (96, 128, 960) to demonstrate our method's effectiveness. 
For multi-GPU experiments, we selected larger datasets, scaling up to 50M in size and 768 in dimension.
We took the first 10 million and the first 50 million part to create the Deep-10M and Deep-50M datasets from the Deep-1B dataset~\cite{deep-1b}.
For graph building of \thiswork, we used CAGRA's graph build algorithm, which offers the fastest build speed on GPUs.
To ensure fairness, the out-degree of graphs was fixed to 64 for all datasets for \thiswork's search and the CAGRA baseline evaluation. 


\textbf{Query Batch Size.}
We evaluated the performance and accuracy of \thiswork and baselines using a query batch size of 10,000 for single-GPU tests.
The exception was Gist-1M, where we used a batch size of 1,000.
For multi-GPU evaluations, we employed a batch size of 60,000 to achieve high throughput and fully utilize multiple devices.

\textbf{Performance Metrics.} 
We tested two key metrics for the evaluation, following the typical performance metrics in the previous ANNS frameworks. The two key metrics and the details are as follows:

\begin{enumerate}
    \item \textbf{Recall}: This metric quantifies the accuracy of the ANNS results by comparing them to the ground truth k-NNS results. For a query $q$, recall is defined as:
    \begin{equation}
        Recall@k = \frac{|\mathcal{N}_k^{KNNS}(q) \cap \mathcal{N}_k^{ANNS}(q)|}{|\mathcal{N}_k^{KNNS}(q)|},
    \end{equation}
    where $\mathcal{N}_k^{KNNS}(q)$ represents the $k$-NNS result, and $\mathcal{N}_k^{ANNS}(q)$ is the result obtained from the ANNS framework.
    Similar to prior work~\cite{ggnn, cagra, song, bang}, we target 95\% recall@10 for most of the analyses unless otherwise noted.
    
    \item \textbf{Queries Per Second (QPS)}: This metric captures the throughput of the framework by measuring the number of queries processed per second.
\end{enumerate}

The primary objective of ANNS frameworks is to balance these two metrics---achieving high recall to ensure accuracy while maximizing QPS to deliver fast query processing.

\subsection{Evaluation on Multiple GPUs}

\begin{figure*}[t]
    \centering
    \includegraphics[width=.97\textwidth]{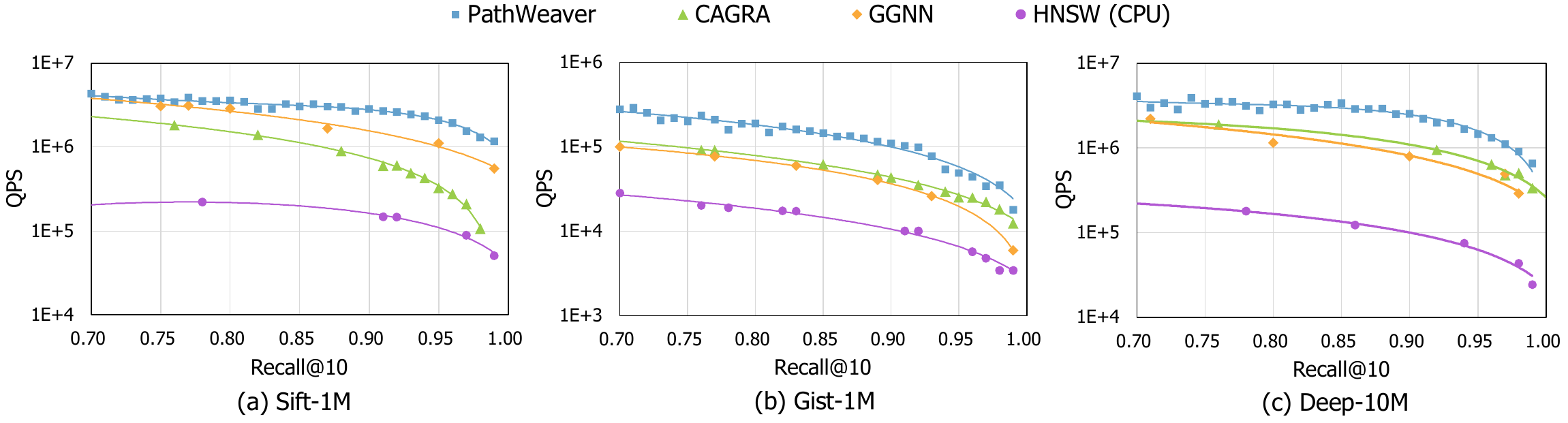}
    \caption{Performance comparison on a single GPU.
    }
    \label{fig:eval:single}
\end{figure*}

\subsubsection{Performance Comparison on Multiple GPUs}
\label{sec:eval:multi}

\cref{fig:eval:multi} presents the performance comparison of \thiswork with other multi-GPU baselines in a QPS-recall plot.
Because CAGRA~\cite{cagra} does not support a multi-GPU environment out-of-box, we extended it with the sharding method as discussed by the authors.
As shown in the results, \thiswork outperforms all baselines. 
At a high recall of 95\%, \thiswork achieves 3.24$\times$ geomean speedup over CAGRA, the best-performing baseline. 
Moreover, \thiswork achieves at most $5.30\times$ speedup at the same recall rate in the Wiki-10M dataset.
A similar advantage is maintained at a moderate recall 88\%--92\%, where \thiswork achieves 3.36$\times$ speedup over CAGRA.

Among datasets, Deep-10M and Deep-50M both show similar QPS of around $3\times10^6$ for \thiswork and $1.1\times10^6$ for both CAGRA and GGNN. 
However, for Wiki-10M, the throughput is an order of magnitude lower in all frameworks because Wiki-10M is composed of very wide vector indices of 768.
This aligns with our finding that the ANNS throughput is less impacted by the size of the graphs. 
Even though Deep-50M has five times more vertices than that of Deep-10M, the number of iterations until convergence is similar, which supports the motivation for \cp.
However, having a wider vector per vertex in Wiki-10M dataset directly affects the performance. 
In this case, \thiswork achieves more speedup than the other two datasets mainly due to \dgp.



To demonstrate the efficiency and scalability of \thiswork, we evaluated its performance while increasing the number of GPUs from 1 to 4, at a recall of 95\%.
The results are shown in \cref{fig:eval:multi_scale}a.
Compared to that of using a single GPU, using four GPUs achieves 2.47$\times$ more speedup when comparing the fastest cases of each GPU setting, which represents 62\% scale efficiency.
Compared to that of the baselines shown in \cref{fig:cp_moti} for the same dataset Deep-10M, the scale efficiency is improved by 17\%.

\cref{fig:eval:multi_scale}b further evaluates the impact of \cp on multiple GPUs. 
Compared to \thiswork using the sharding method of baseline (denoted as `\Naive \thiswork'), the advantage of \cp is maintained across various datasets and target recall values.
This indicates that \thiswork scales stably on various settings.

\subsection{Performance Comparison on a Single GPU}

While \thiswork provides a significant speedup with multiple GPUs, its benefit remains in single-GPU settings. 
Except for \cp, \thiswork can still benefit from \sr and \dgp on a single-GPU setting. 
We plotted the performance of \thiswork and other baselines in \cref{fig:eval:single}.
In addition to the GPU-based methods, we additionally study HNSW~\cite{hnsw}, a baseline method implemented on a CPU.

Overall, \thiswork achieves a speedup of 3.43$\times$ over CAGRA.
This speedup is primarily attributed to reduced distance computations due to \dgp and fewer iterations resulting from the application of \sr.
Similar to the observations from multi-GPU experiments, Gist-1M results in the slowest throughput due to its larger vector dimension of 960.
In all cases under evaluation, \thiswork exhibited better QPS-recall trade-off compared to the baselines.

\begin{figure}[t]
    \centering
    \includegraphics[width=.99\columnwidth]{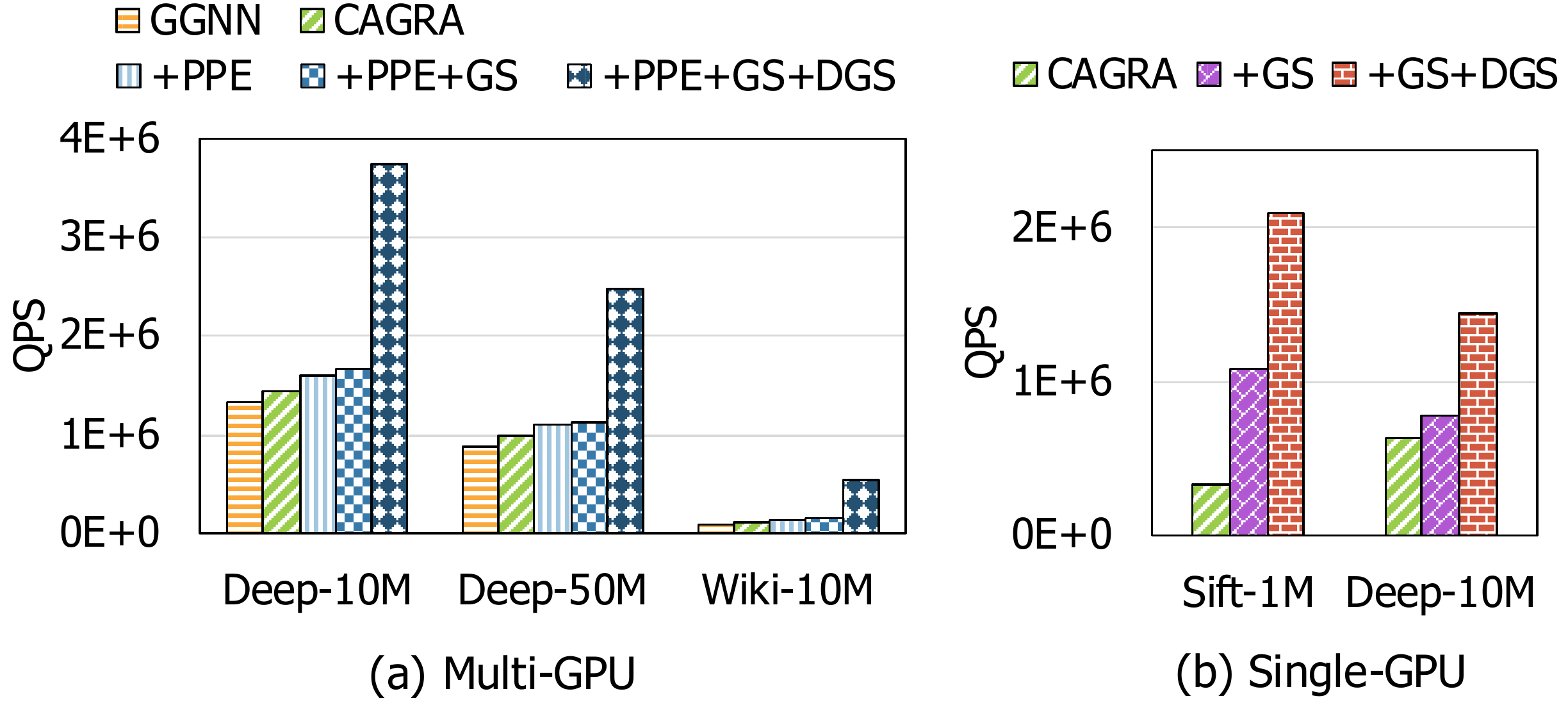}
    \caption{Ablation study of \thiswork on (a) multi-GPU and (b) single-GPU settings. `+PPE', `+GS', and `+DGS' represent \thiswork with \cp, \sr, and \dgp, respectively.
}
    \label{fig:eval:abl}
\end{figure}

\subsection{Ablation Study}

To demonstrate the impact of each scheme in \thiswork, we conducted an ablation study using the Deep-10M, Deep-50M, and Sift-1M datasets on four GPUs, as well as Deep-10M and Sift-1M on a single GPU.
In the results shown in \cref{fig:eval:abl}, PPE, GS, and DGS represent \cp, \sr, and \dgp, respectively.


For the ablation study on multiple GPUs, we used GGNN and CAGRA as baselines, and gradually applied PPE, GS, and DGS to analyze the effect of each, as shown in \cref{fig:eval:abl}a.
On top of the baselines, each component of \thiswork adds consistent speedup across datasets with similar trends.
In addition, \cref{fig:eval:abl}b shows the ablation study on the single-GPU setting.
Although \cp cannot be applied in this case, \sr provides higher speedup.
This is because in the single-GPU setting, the entire search can be regarded as the first stage, which benefits from the ghost stage.



\subsection{Execution Time Breakdown}
\label{sec:eval:multi_break}



\begin{figure}[t]
    \centering
    \includegraphics[width=.97\columnwidth]{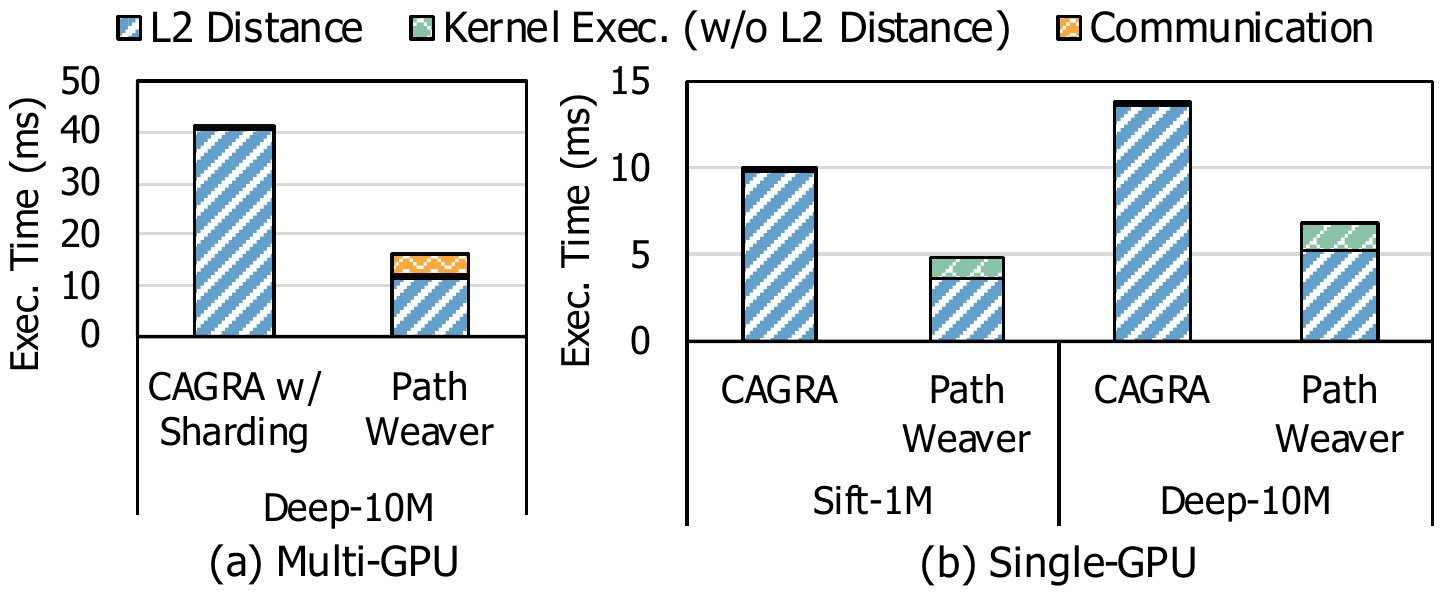}
    \caption{Search execution time breakdown on (a) multi-GPU and (b) single-GPU setting.
    }
    \label{fig:eval:breakdown}
\end{figure}

To investigate how the execution time is spent in \thiswork, we further broke down the execution time into three components: L2 distance calculation, rest of the kernel execution time, and inter-GPU communication time.
The rest of the kernel includes random number generation, neighbor fetching, distance sorting, and hash table management for avoiding duplicate node visits.
They are evaluated at 95\% recall, with Deep-10M dataset for multi-GPU setting and an additional Sift-1M dataset for the single-GPU setting.
To obtain the breakdown, we used \texttt{clock64()} function to separate the portion of L2 distance calculation within a kernel. 
For breakdowns of multi-GPU execution time, we measured the execution time with a certain portion of the code disabled, similar to the CPI stack~\cite{cpistack} method.

The multi-GPU results are shown in \cref{fig:eval:breakdown}a.
We compare \thiswork with extended CAGRA implementation for multi-GPUs (CAGRA w$\slash$ Sharding).
As shown in the result, the L2 distance calculation is the dominating factor in both the CAGRA w$\slash$ Sharding and the fully optimized \thiswork. 
In CAGRA w$\slash$ Sharding, no execution time is attributed to inter-GPU communication, since all searches are performed independently without any data exchange across GPUs.
In \thiswork, the communication time is incurred because of inter-stage data transfer of \cp. 
However, this results in fewer iterations required until convergence in subsequent stages, leading to a smaller L2 distance calculation.
The rest of the kernel's portion slightly increases with optimizations in \thiswork, due to the additional direction flag lookup process introduced by \dgp. 
However, its portion is almost negligible in both CAGRA w$\slash$ Sharding and \thiswork, making its impact minimal.

\cref{fig:eval:breakdown}b depicts the breakdown on a single GPU.
Because there is no communication, most of the time is spent on the L2 distance calculation. Thus, the speedups come from \sr and \dgp.
Instead of communication, the miscellaneous kernel execution time slightly increases in \thiswork, which accounts for the ghost stages and the reduction of the dominant L2 distance calculation.



\subsection{Detailed Analysis}

In this subsection, we further investigate the effect of \thiswork in terms of \cp, \sr, and \dgp.

\subsubsection{\CP}

\begin{figure}[t]
    \centering
    \includegraphics[width=.99\columnwidth]{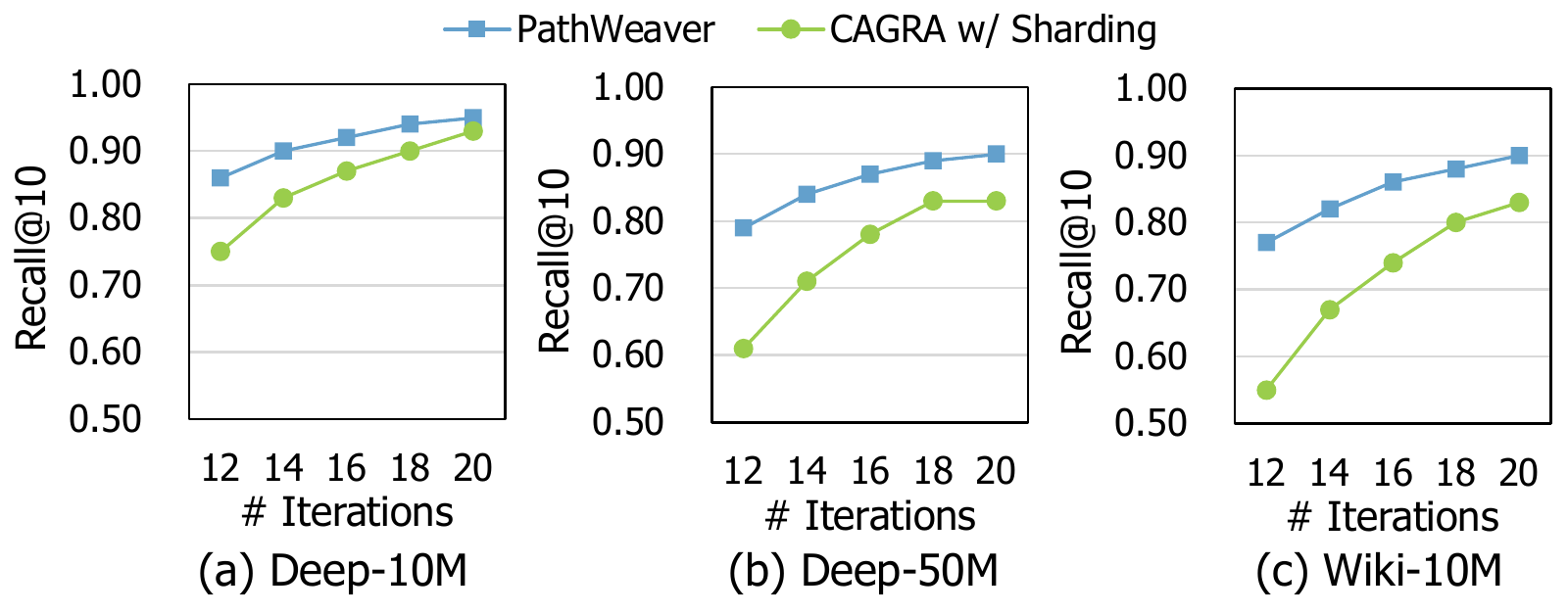}
    \caption{Accuracy comparison on different number of iterations.
    }
    \label{fig:eval:cp_detail}
\end{figure}

To demonstrate the effect of \cp, we analyzed how the recall rate evolves by increasing the max number of search iterations in \cref{fig:eval:cp_detail} on three datasets (Deep-10M, Deep-50M, and Wiki-10M).
\thiswork achieves high recall values with significantly fewer iterations.
This is due to \cp, which enables each GPU to initiate the search process from a data point closer to the query using the search results from other GPUs.
In contrast, the baseline requires substantially more iterations to reach comparable recall rates. 
For example, in the Deep-10M dataset, the baseline reaches a recall rate 0.90 with 18 iterations, while \thiswork achieves this with only 14 iterations.  

\subsubsection{\SR} 

\begin{figure}[t]
    \centering
    \includegraphics[width=.97\columnwidth]{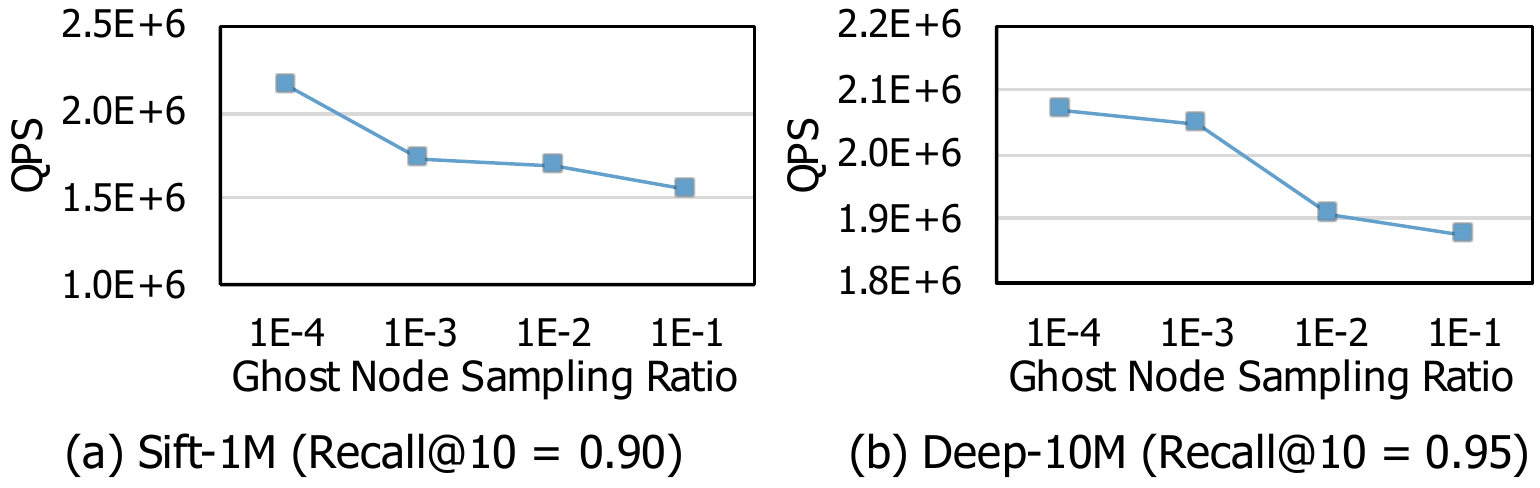}
    \caption{Sensitivity study on the relationship between the ghost node sampling ratio and QPS.
    }
    \label{fig:eval:sr_detail}
\end{figure}

To evaluate the impact of the sampling ratio of ghost nodes on search performance, we conducted search operations on the Sift-1M and Deep-10M datasets using a single GPU, while varying the sampling ratio for building the ghost shard. 
As shown in \cref{fig:eval:sr_detail}, higher QPS was observed at lower sampling ratios of ghost nodes.
For example, the QPS is $1.39\times$ higher for sampling ratio $0.0001$ compared to $0.1$ in the Sift-1M dataset. 
We attribute this to the connections among the smaller set of ghost nodes, which facilitate larger iteration jumps within the original graph. 
The result indicates that a small-sized ghost shard is often sufficient, which provides another explanation for why \sr can bring speedup. 


\subsubsection{\DGP} 

\begin{figure}[t]
    \centering
    \includegraphics[width=.99\columnwidth]{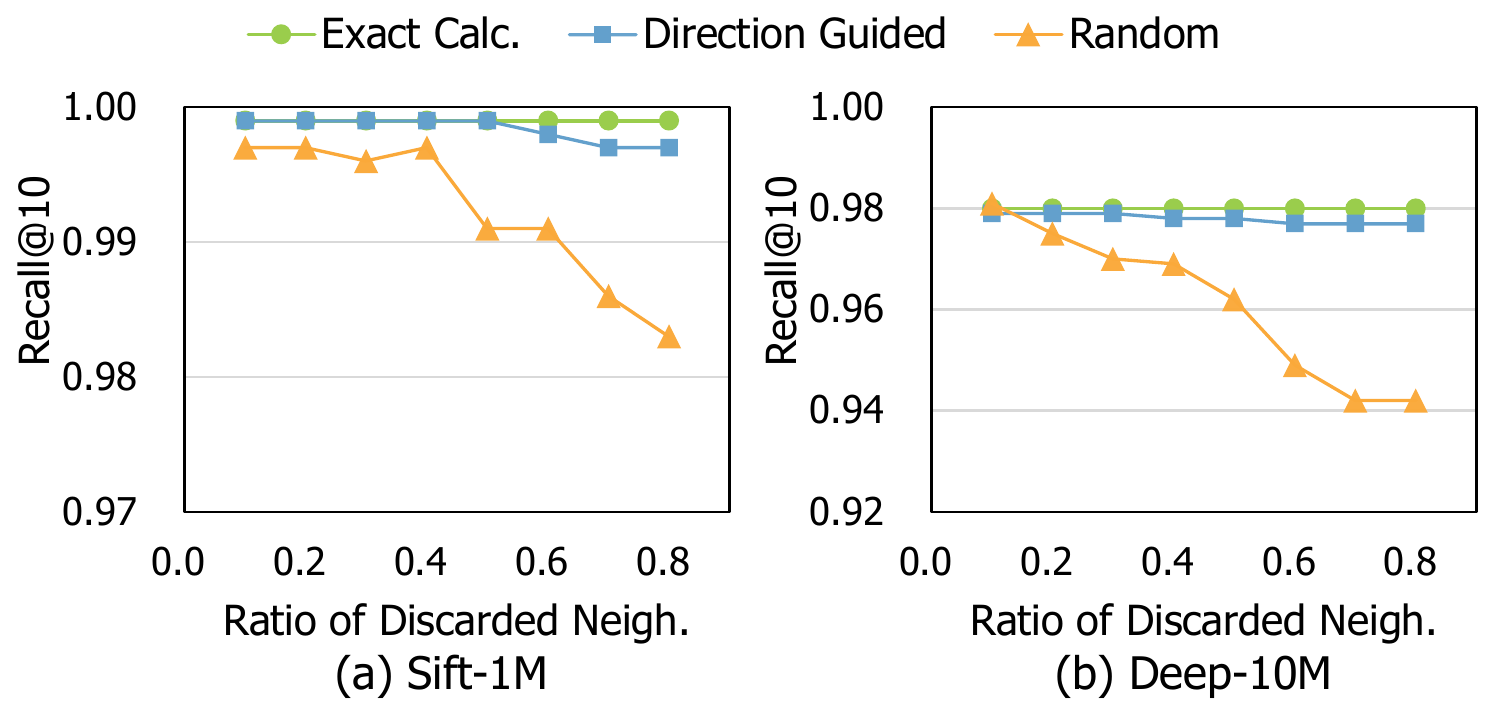}
    \caption{Comparison of neighbor selection strategies by varying the ratio of discarded neighbors.
    }
    \label{fig:eval:dgp_detail_neigh}
\end{figure}
\begin{figure}[t]
    \centering
    \includegraphics[width=.99\columnwidth]{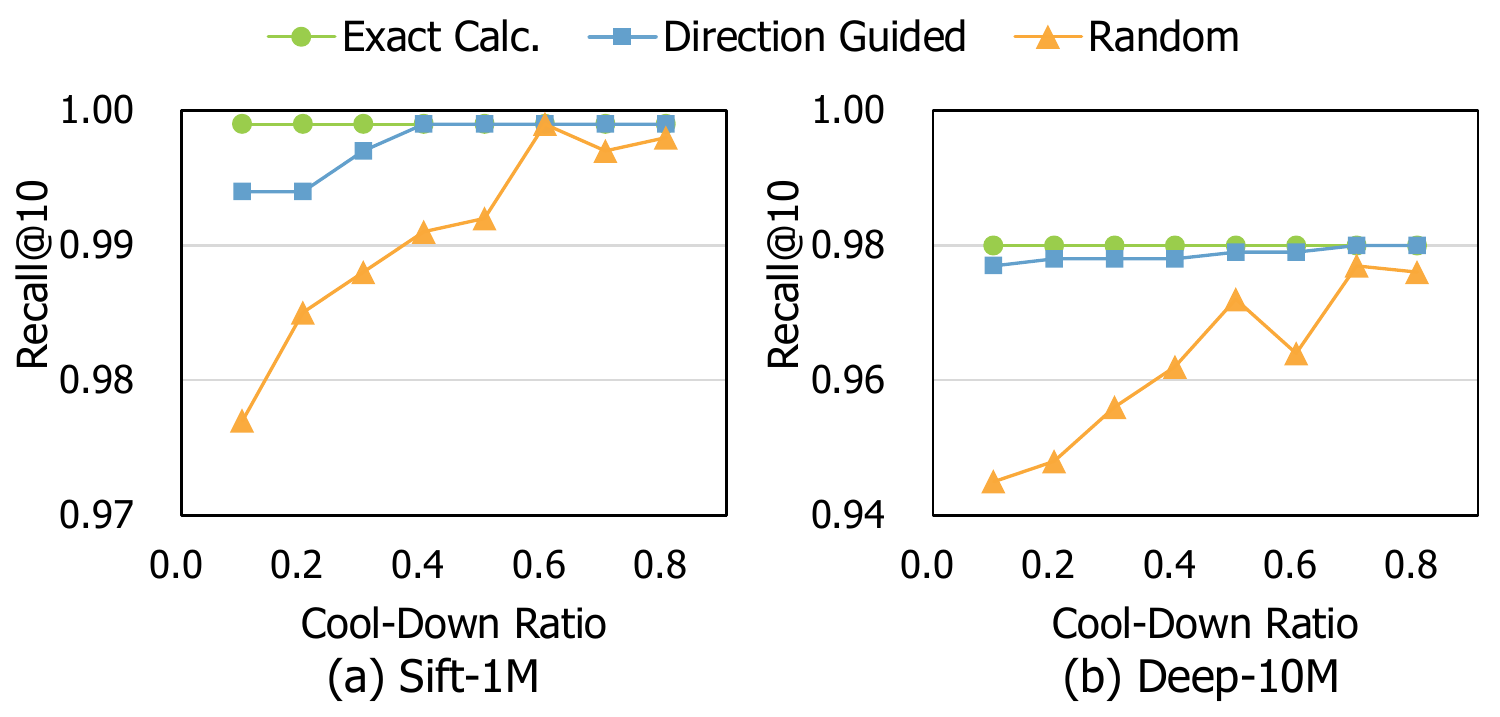}
    \caption{Comparison of neighbor selection strategies by varying the cool-down ratio of search iterations against the total iteration. A small cool-down ratio indicates that \dgp or random selection is applied during the majority of the iterations.
    }
    \label{fig:eval:dgp_detail_iter}
\end{figure}

To demonstrate the effectiveness of using direction bits as the neighbor discarding metric in \dgp, we compared the performance of \dgp against exact calculation (no discarding) and random neighbor discarding (randomly selecting neighbors).

Initially, the recall rate was evaluated for various discarded neighbor ratios while using a cool-down ratio of 0.5, as shown in \cref{fig:eval:dgp_detail_neigh}.
Compared to exact calculations, which achieve ideal recall, neighbor discarding in \dgp results in a slight recall drop of at most 0.003, whereas random neighbor discarding causes a significant recall degradation of at most 0.038 for the Deep-10M dataset.
Notably, \thiswork maintains robust recall even with a discarding ratio of 0.7, in contrast to the significant recall decline observed with the random method.

Next, we analyze the recall rate by adjusting cool-down ratios with a fixed neighbor discarding ratio at 0.5, as shown in \cref{fig:eval:dgp_detail_iter}.
Similar to the prior analysis, \thiswork shows robust recall even when the cool-down ratio decreases, while the random method fails to maintain high recall. 

For example, at a cool-down ratio of 0.3, neighbor discarding in \dgp demonstrates only a minor recall drop of 0.002, compared to a significant degradation of 0.032 with the random discarding approach on the Deep-10M dataset.
Both analyses demonstrate the effectiveness of using the direction of vectors when filtering out neighbors.

\subsection{Graph Build Time Analysis}
\label{sec:eval:build}

\begin{figure}[t]
    \centering
    \includegraphics[width=.99\columnwidth]{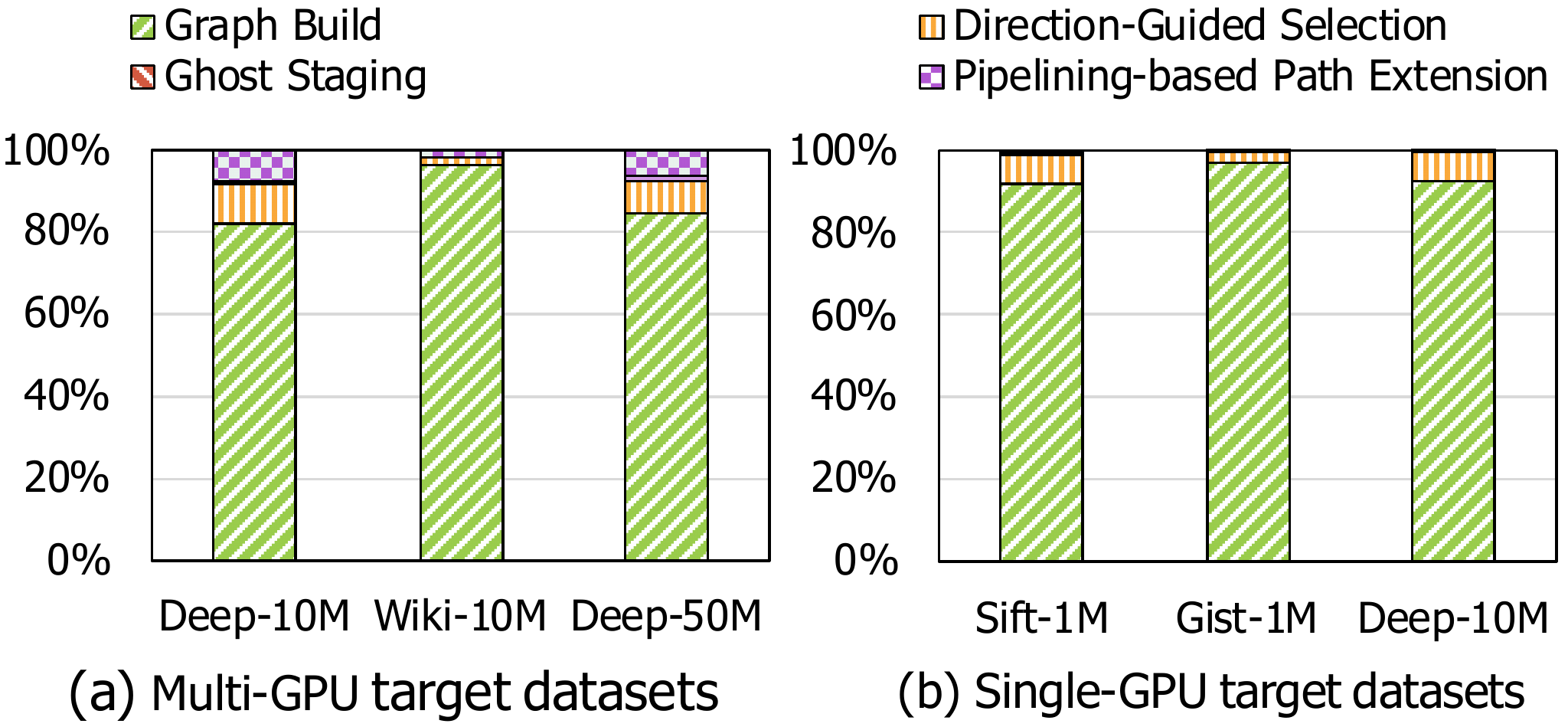}
    \caption{Graph build time overhead analysis. 
    }
    \label{fig:eval:build}
\end{figure}

The preprocessing steps of \thiswork, such as generating inter-shard connection edges for \cp, ghost nodes connections for \sr, and direction bit vectors for \dgp, necessitate an overhead analysis to evaluate their impact.
Therefore, we perform a breakdown analysis of the proximity graph build time, as shown in \cref{fig:eval:build}.
Note that \thiswork uses the graph build algorithm of CAGRA~\cite{cagra}, represented as `graph build'. 

The results indicate that the overall overhead is less than 10\% for datasets targeting a single GPU.
The overhead of constructing ghost node connections was lower than generating direction bit vectors.
This is because the latter requires creating direction bit vectors for all edges in the graph, while the former requires connecting a relatively smaller number of data points. 
For multi-GPU target datasets, the overhead was less than 4\% for Wiki-10M, while it reached 15\% for Deep-50M. 
This is because building the graph also involves many L2 distance calculations, whose overhead is proportional to the vector dimension, similar to the search kernel.
Additionally, the overhead of constructing ghost shards remains negligible, with the overheads of generating inter-shard connections and direction bit vectors being comparable.
The minimal overhead of creating inter-shard connections arises from leveraging the existing connections in the proximity graph. 
These connections provide information about which node in the next shard is closest to the target node in the current shard.
Overall, for both cases of single- and multi-GPU settings, the additional graph build overhead from \thiswork's designs is small. 

\section{Discussion}

\begin{figure}[t]
    \centering
    \shepindex{S1}
    \fcolorbox{olivegreen}{white}{
        \includegraphics[width=.99\columnwidth]{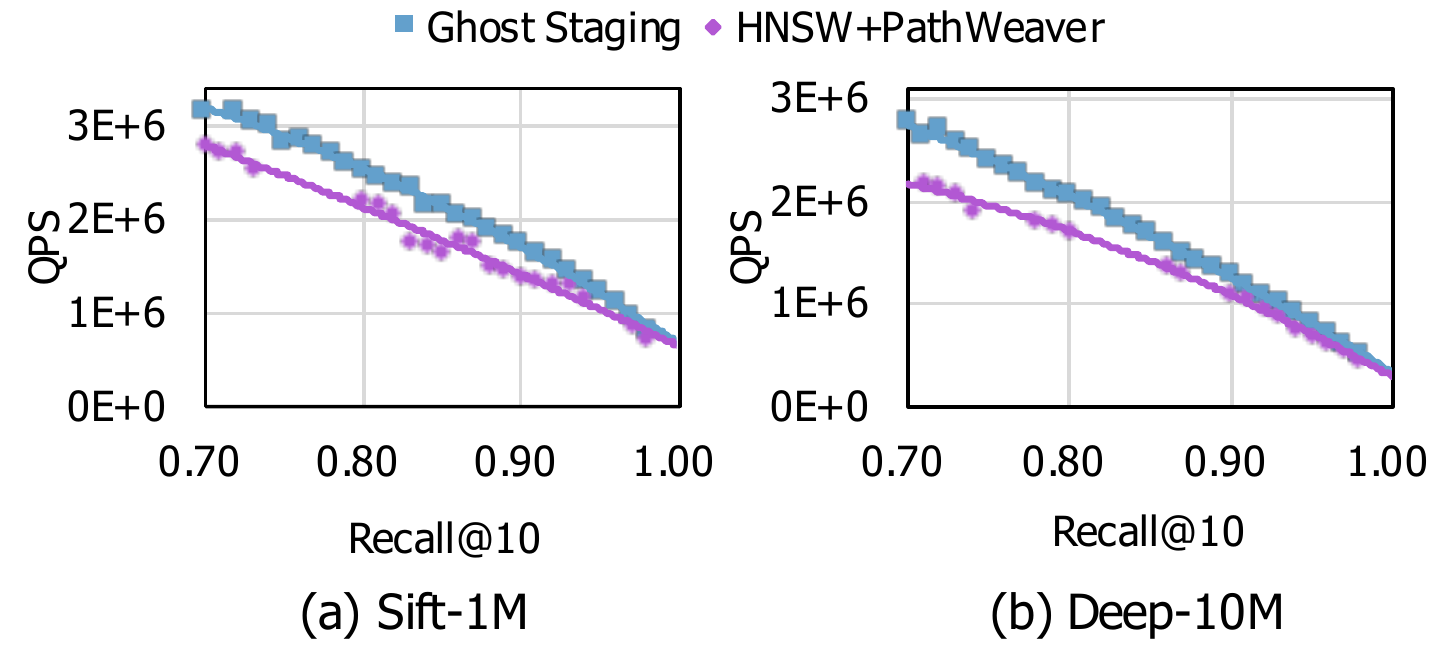}
    }
    \caption{Comparing \sr and GPU-based HNSW.
    }
    \label{fig:disc:gs_hnsw}
\end{figure}

\shep{
\subsection{Comparing \SR and Existing Hierarchical Graph Approaches}
\shepindex{S1}
Although \sr is a kind of hierarchical method, it differs from existing methods such as HNSW~\cite{hnsw} and GGNN~\cite{ggnn} in its objective and implementation.
HNSW sequentially inserts each node into level $n$ with exponentially decreasing probability as $n$ increases, constructing a proximity graph at each layer.
On the other hand, GGNN partitions the nodes, builds a graph within each partition, and finally merges them by selecting nodes for the next layer.

In contrast, \sr starts from an already-built proximity graph, and the extra layer is constructed by sampling vertices and connecting them.
While existing hierarchical solutions aim to improve overall reachability and convexity by introducing hierarchy during graph construction, our approach uses the additional ghost stage solely to efficiently identify an entry node on the base proximity graph.

To investigate the difference, we conduct searching on the HNSW graph with our GPU-based search kernel and compare the results with those of \sr (\dgp and \cp are disabled for fair comparison).
As depicted in \cref{fig:disc:gs_hnsw}, \sr consistently achieves faster search speed compared to the HNSW graph even when it's searched with GPUs. 
}

\shep{
\subsection{Managing Dynamic Updates}
\shepindex{S4}

While \thiswork currently targets static graphs, it is worthwhile to discuss dynamic updates to the proximity graphs~\cite{xu2022proximity}.
Because \thiswork is based on shards, any update would only involve modifying the affected shard. 
When rebuilding is needed, the cost mostly comes from the affected local graph, because the auxiliary data introduced by \thiswork only accounts for a small portion as reported in \cref{fig:eval:build}. 


For a small number of insertions, nodes can be added to one of the existing shards, followed by a rebuilding for the associated graph and the auxiliary data.
If the number of insertions is small enough, the growth of the shard will not cause load imbalance.
In such a scenario, the inter-shard edges can be incrementally updated because the small change in the local graph does not affect the similarity between existing vertices.
For a large number of insertions, an additional shard can be created without modifying existing ones. 
For a small number of deletions, a deleted node may still act as a bridge between its neighbors.
To preserve connectivity, a deletion flag can be used to logically remove the node. 
However, when a substantial portion of a shard is deleted, rebuilding the shard and its associated structures becomes beneficial.
}

\shep{
\subsection{Comparing \DGP with Static Graph Pruning Strategy}
\shepindex{S5}
In ANNS, several approaches perform graph pruning~\cite{cagra,diskann} in a static manner to reduce memory usage and improve query efficiency.
This strategy reduces graph density by eliminating less important edges during index construction.
While static pruning offers a consistent structure across all queries, it may not be optimal for query-specific behaviors.

On the contrary, \dgp can be considered a dynamic pruning strategy.
Instead of permanently discarding edges, 
it dynamically chooses a subset of neighbors during the search process, based on the direction to the query.
One potential issue of its fixed top-$n$ selection is that it might discard important candidates. 
Even though we did not observe a significant drop in recall in our experiments, one could instead prune based on the similarity criteria, as done in \cite{zhao2024guitar} for neural ranking.
Such a method could preserve good candidates, at the potential cost of warp imbalances due to non-uniform pruning.
}

\shep{
\subsection{Overhead Analysis of \CP}
\label{sec:disc:pp_cost}
\shepindex{S2}
In this subsection, we quantify the communication overhead of \cp.
First, the volume of inter-GPU communication is only $Q \times b_{idx}$, where $Q$ is the number of queries and $b_{idx}$ is the number of bytes per transmitted index. 
In contrast, the amount of GPU memory accesses scales with $I \times J \times Q \times v \times b_{elem}$, where $I$ is the number of search iterations, $J$ is the out-degree of each node, $v$ is the vector dimension, and $b_{elem}$ is the number of bytes per vector element.

Although inter-GPU channels (e.g., NVLink) are typically over $10\times$ slower than GPU's memory bandwidth, the dominant cost still lies in the GPU's memory term, as the product $J \times v$ often exceeds $10^4$. 
As a result, the reduction in $I$ achieved by \cp has a much greater impact on total latency than the relatively small inter-GPU communication cost. 
}



\section{Related Works}

\subsection{Graph-based ANNS Solutions}

\textbf{CPU-based solutions.}
A range of CPU-based solutions has been proposed to enhance the construction and traversal of proximity graphs for approximate nearest neighbor (ANN) search.
Key studies, such as \cite{nsw, hnsw, efanna, nsg, hcnng}, focus on designing efficient graph structures that are leveraged by the beam search algorithm to achieve scalable and effective ANN retrieval.
To improve search efficiency, several methods incorporate further optimizations~\cite{grasp, graphreorder_cache, finger, quickadc, adsampling, pecann, parlayann, highdim}.
For example, subgraph sampling and edge pruning techniques are utilized in \cite{grasp} to reduce traversal overhead, while \cite{graphreorder_cache} explores reordering graph structures to enhance cache efficiency.
Additionally, \cite{finger} introduces a novel distance function approximation that accelerates the search process, demonstrating significant performance gains in graph-based ANNS.

\textbf{GPU-based solutions.}
SONG~\cite{song} represents the first graph-based ANN search solution for GPUs, leveraging GPU parallelism to address the bottleneck of computationally expensive distance calculations while optimizing data structures through various approaches.
GANNS~\cite{ganns} and GGNN~\cite{ggnn} build upon this by focusing on GPU-friendly implementations that efficiently utilize shared memory for maintaining data structures and parallelizing their operations.
CAGRA~\cite{cagra} enhances graph construction and search processes on GPUs, employing techniques such as warp splitting and forgettable hash to achieve high throughput.
On the other hand, \thiswork moves beyond the sole reliance on proximity graphs during the search process.
By integrating information from adjacent shards and considering query directionality, it achieves enhancements in both speed and accuracy. 

\textbf{Other platforms.}
Graph-based ANNS solutions have also been explored on alternative platforms to enhance performance~\cite{pyramid, csdanns1, smartanns, ndsearch, proxima, df-gas, cxl-anns}.
Pyramid~\cite{pyramid} introduces a hierarchical near-memory-computing (NMC) architecture specifically designed for efficient graph-based ANNS. 
Computational storage devices (CSD) are leveraged in works such as~\cite{csdanns1, smartanns}, where software-hardware co-design accelerates search algorithms by integrating computation closer to data storage. 
DF-GAS~\cite{df-gas} proposes a distributed FPGA-as-a-Service architecture, enabling parallel search operations across both full graphs and subgraphs.
CXL-ANNS~\cite{cxl-anns} employs memory disaggregation from the host via CXL to optimize graph-based ANNS performance.

\subsection{Scaling ANNS to Large Datasets}
For ANNS on large datasets, various approaches focus on compressing data vectors. 
Quantization-based methods~\cite{pq, opq, lopq, vecquant} achieve this by clustering nodes within a graph and replacing them with centroid-based representations. 
Similarly, SONG~\cite{song} utilizes 1-bit random projection to enable large datasets to fit within GPU memory.

Another prominent approach to handling large datasets involves hierarchical or hybrid architectures. 
Methods such as \cite{diskann, hmann, filtered-diskann, ood-diskann, lm-diskann, freshdiskann} utilize storage devices to store large datasets, whereas BANG~\cite{bang} combines CPU and GPU resources by maintaining the large graph index in CPU memory and leveraging GPU acceleration. 
FusionANNS~\cite{fusionanns} employs a combination of CPU-GPU collaboration and SSDs to optimize performance.
These approaches enable cost-efficient processing of large datasets, though the achievable speedup remains inherently limited by the constraints of single-machine architectures.

Sharding datasets to fit into device memory and utilizing multiple devices has been widely investigated.
Methods such as \cite{ggnn, billscalegpu, CHEN2019295, csdanns1} employ a multi-device strategy by partitioning datasets across devices, processing queries independently on each, and aggregating the results on the host.
While these approaches enable distributed processing, achieving substantial and consistent speed-up as the number of devices increases often remains challenging.
In contrast, our \cp achieves a proportional speedup with the number of machines used, effectively addressing the scalability limitations observed in previous methods.

\section{Conclusion}

We propose \thiswork, a multi-GPU framework for graph-based ANNS that provides scalable and efficient performance on large datasets. 
\thiswork addresses the main bottlenecks of graph-based ANNS through three key innovations: GPU-to-GPU communication to reduce unnecessary search iterations, a representative dataset search to refine starting data points, and data point filtering to minimize unnecessary memory access and distance calculation. 
Our evaluation results demonstrate that \thiswork outperforms both CPU- and GPU-based solutions, achieving higher performance while maintaining accuracy.  

\section*{Availability}

We have open-sourced the code of \thiswork and made it publicly available. The GitHub repository for \thiswork can be found at \url{https://github.com/AIS-SNU/PathWeaver}.

\section*{Acknowledgement}

This work was partially supported by 
National Research Foundation of Korea (NRF) grant funded by the Korea government (MSIT) (2022R1C1C1011307), 
and
Institute of Information \& communications Technology Planning \& Evaluation (IITP) (RS-2024-00395134, 
RS-2024-00347394, 
RS-2023-00256081, 
RS-2021II211343).   
Jinho Lee is the corresponding author.

\bibliographystyle{plain}
\bibliography{references}

\end{document}